\documentclass[prl,a4paper,superscriptaddress,twocolumn,amsmath,amssymb,floatfix,amstex]{revtex4-1}%
\usepackage{graphicx,graphics,rotating}	% Include figure files
\usepackage{dcolumn}
\usepackage{bm}
\usepackage{hyperref}
\usepackage{color}
 \usepackage{soul}   								%%% package that includes \st command to "strikethrough" words
\begin{document}
\title{Glassy properties of Anderson localization: pinning, avalanches and chaos. }

\author{G.~Lemari\'e}
\email[Corresponding author: ]{lemarie@irsamc.ups-tlse.fr}
\affiliation{%
Laboratoire de Physique Th\'eorique, IRSAMC, Universit\'e de Toulouse, CNRS, UPS, France
}

\date{September 6, 2018}% 
%%%

\begin{abstract} {I present the results of extensive numerical simulations which reveal the glassy properties of Anderson localization in dimension two at zero temperature: pinning, avalanches and chaos. I first show that {strong localization confines quantum transport along paths which are pinned by disorder but can change abruptly and suddenly (avalanches) when the energy is varied.} I determine the roughness exponent $\zeta$ characterizing {the transverse fluctuations of these} paths and find that {its value $\zeta=2/3$ is the same as for the} directed polymer problem. {Finally, I characterize the chaos property, namely the
fragility of the conductance with respect to small perturbations in
the disorder configuration. It is linked to
interference effects and universal conductance fluctuations at weak
disorder, and more spin-glass-like behavior at strong disorder.}}  
\end{abstract}

\maketitle

\emph{Introduction.---}
Anderson localization (AL) \cite{Anderson:PR58, abrahams201050} is a key mechanism of non-ergodicity in disordered \emph{quantum} systems, with prominent examples such as the insulating state of disordered materials \cite{dobrosavljevic2012conductor}, quantum multifractality  {\cite{castellani1986multifractal, wegner1980inverse, evers2008anderson, feigel2010fractal}} or the absence of thermalization for many-body closed systems {\cite{PhysRevLett.78.2803, basko2006metal, PhysRevLett.95.206603, nandkishore2015many, abanin2018ergodicity}}. 
Spin glass physics is another paradigm of non-ergodic behavior which arises in \emph{classical} disordered systems. 
Its study has led to important theoretical breakthroughs like the concept of spontaneous replica symmetry breaking \cite{mezard1987spin} and has found applications in e.g. optimization or biology \cite{young1998spin, stein2013spin}.

There have been only few analogies drawn between these two domains. In strongly localized materials, the electron glass \cite{davies1982electron} has been much discussed in the literature \cite{PhysRevLett.93.256403, PhysRevLett.101.056601, PhysRevLett.84.3402, amir2011electron}. This glassy phase arises {at finite temperature} in the hopping regime \cite{pollak1991hopping} where transport is mediated by phonons. Recently, the Anderson transition on random graphs {of effective infinite dimensionality} has also raised a strong interest \cite{monthus2008anderson, biroli2012difference, deluca2014anderson, kravtsov2015random, tikhonov2016anderson, PhysRevLett.118.166801, biroli2017delocalized}. There is now a consensus that the delocalized phase on trees is non-ergodic \cite{tikhonov2016fractality, facoetti2016non, kravtsov2017non, sonner2017multifractality}, a property related to replica symmetry breaking \cite{kravtsov2017non}.
{In this article, I address the intrinsic glassy properties of AL at zero temperature, i.e. in a fully coherent regime distinct from the electron glass problem. Moreover, I deal with a finite dimensional case $d=2$, more realistic than the random graph case.} 

{In the strongly localized regime, quantum transport has been conjectured to be strongly inhomogeneous and, in particular, to follow directed paths \cite{nguyen1985tunnel, fritzsche1990hopping, medina1989interference, PhysRevB.93.054201, PhysRevB.46.9984}}. Because of this, strong AL {has been} believed to be analogous to the directed polymer (DP) problem, one of the simplest statistical physics models for which the disorder plays a quite non-trivial role \cite{halpin1995kinetic}. This analogy was confirmed recently by numerical simulations \cite{prior2005conductance, PhysRevLett.99.116602, prior2009conductance, PhysRevB.91.155413} which showed that the fluctuations of the conductance follow the universal {fluctuation properties of DP. Nevertheless, these fluctuation properties do not reflect all the physics of DP, and in particular its glassy nature \cite{derrida1988polymers, mezard1990glassy}.} This article goes a step further by addressing the {new} glassy properties that Anderson localization could inherit from its similarity with DP physics.

Indeed, in dimension two, it is known that the DP problem is in a glassy phase where it exhibits several characteristic glassy properties: pinning \cite{halpin1995kinetic}, avalanches \cite{mezard1990glassy, PhysRevE.57.6936, thiery2017analytical} and chaos \cite{mckay1982spin, PhysRevLett.58.57, PhysRevB.38.386, PhysRevB.43.10728, sales2002fragility, da2004temperature, PhysRevLett.96.235702}. {Then, a DP is} pinned in a configuration that does not move continuously when the system is smoothly perturbed, but sometimes makes an avalanche, i.e. jumps brutally into a very different configuration. Chaos {is defined as} the extreme fragility of these glassy states: an infinitesimal perturbation induces a complete reorganization of the equilibrium configurations at the thermodynamic limit. Chaos was predicted originally for spin glasses and then it was realized that DP, and more generally elastic objects pinned by disorder \cite{halpin1995kinetic, balents1996large, PhysRevLett.96.235702}, form a kind of ``baby spin glass'' \cite{mezard1990glassy, balents1996large, PhysRevLett.96.235702}. {Such glassy properties have never been discussed in the context of Anderson localization.}

In this letter, I present the results of extensive numerical simulations {which fully take into account the non-trivial interplay between quantum interference and disorder. They reveal that the three glassy properties, pinning, avalanches and chaos, are present in Anderson localization. Chaos is however crucially affected by quantum interference effects.} I have used the recursive Green's function method \cite{datta1997electronic, abbout2011thermal} to access efficiently the zero temperature conductance of many (up to $\approx 7\cdot 10^4$) and large (up to $4 \cdot 10^4$ sites) 2D samples. Moreover, a method \cite{pichard1991quantum, markovs2010electron} similar to scanning gate microscopy \cite{topinka2001coherent, abbout2011thermal, gorini2013theory} allows me to image the directed paths taken by the electron flow in the strongly localized regimes. 
This is complemented by a careful finite-size scaling and droplet scaling arguments. 

\begin{figure}
\includegraphics[height=0.48\linewidth]{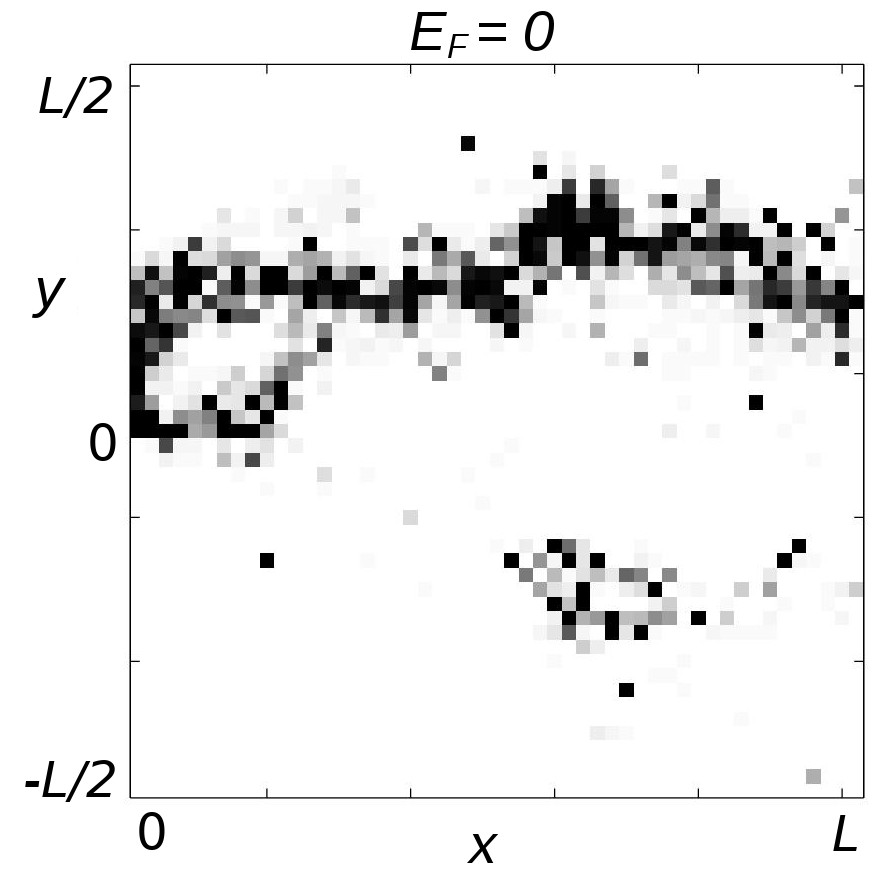}
\includegraphics[height=0.48\linewidth]{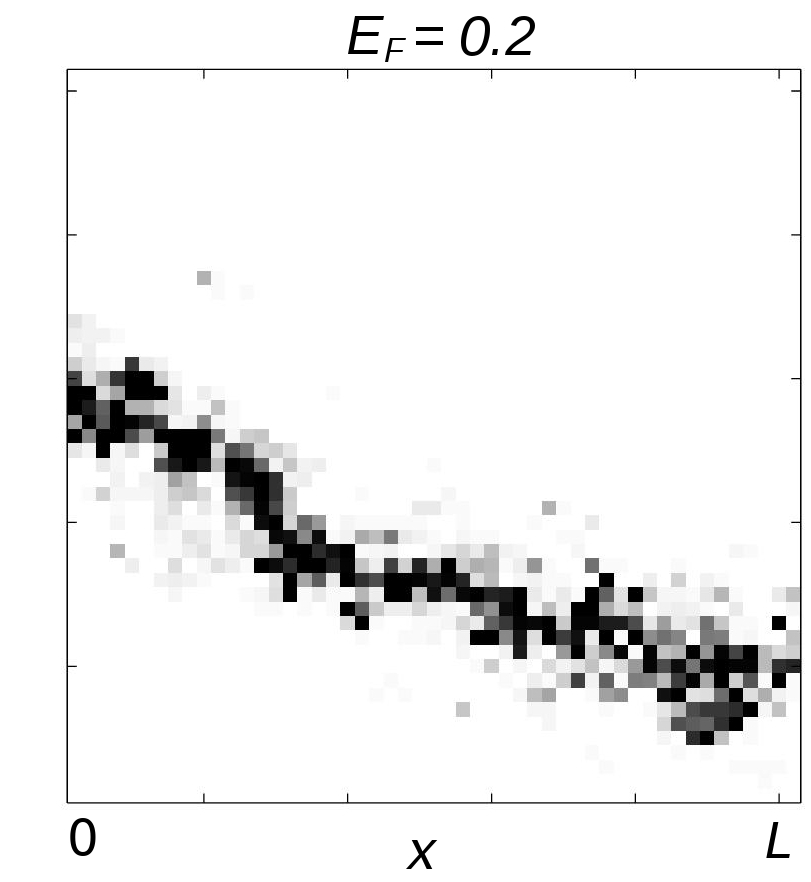}
\caption{In the Anderson localized regime, the zero temperature transport is strongly inhomogeneous and takes place along certain directed paths \cite{markovs2010electron}. Similarly to scanning gate microscopy \cite{topinka2001coherent}, a gray scale plot of the change of conductance $\vert g(i)- g_0 \vert/g_0$, where $g(i)$ is the conductance of the sample with $\varepsilon_i \rightarrow -\varepsilon_i$ and $g_0$ the unperturbed conductance, allows to image the path taken by the electron flow \cite{pichard1991quantum}. A single sample with a box-distributed disorder $W=20$ {($\xi\approx 1.5$)} and size $L=50$ is considered at two values of the  energy $E_F$.}
 \label{fig:AndApath}
\end{figure}

\emph{Zero temperature conductance.---}
I consider the conductance through a scattering system described by an Anderson model \cite{Anderson:PR58} of size $L \times L$:
\begin{equation}
 H = \sum_i \varepsilon_i a_i^\dagger a_i + t \sum_{<i,j>} a_j^\dagger a_i + H.c. \; ,
\end{equation}
where $a_i$ ($a_i^\dagger$) is the annihilation (creation, resp.) operator of an electron at site $i$ of a square lattice, $t=1$ is the hopping amplitude and the sum is restricted to nearest neighbors{, with open boundary conditions along the transverse direction $y$}. The site energies are {independent} random variables having, unless stated, a normal distribution with {zero} mean and standard deviation $W$. 
Such a disordered scatterer is attached to two perfect leads, which can be either wide with the same section of the sample, or narrow, consisting in a 1D lead attached at the middle {($y=0$)} of one edge {($y \in [-L/2,L/2]$, see Fig.~\ref{fig:AndApath} and \cite{supmat})}. The results presented in this letter all correspond to the case of a narrow left lead and a wide right lead, but I have checked that two wide leads have qualitatively similar properties {\cite{supmat}}.

The {zero temperature} dimensionless conductance {at energy $E_F$} of this model is computed through the Fisher-Lee formula \cite{fisher1981relation}
$ g(E_F)~=~\text{Tr}[\Gamma_\mathcal{L} \mathcal G^r \Gamma_\mathcal{R} \mathcal G^a ]$,
which uses the Green's function $\mathcal G(E_F)$ of the scatterer dressed by the leads, with the exponent $r$ ($a$) denoting retarded (advanced, resp.), and $\Gamma_{L,R}=-2 \text{Im}[\Sigma_{L,R}^r]$ the imaginary part of the self-energies associated to the leads, see \cite{datta1997electronic}. The Green's function is efficiently calculated numerically using the recursive Green's function approach \cite{datta1997electronic}.

\emph{Analogy with the directed polymer problem.---}
{Before describing the analysis of the glassy properties of {quantum} transport, I briefly review the {known arguments and the recent numerical studies supporting the analogy with the DP problem.}}
The Green's function element between a point $A$ at the left edge of the scatterer and another point $B$ at the right edge can be expressed formally using the locator expansion \cite{Anderson:PR58}:
$\mathcal G_{A,B}(E_F)~=~\sum_{\mathcal P} \prod_{i\in\mathcal P} (E_F-\varepsilon_i)^{-1}$, where the sum is over all paths $\mathcal P$ connecting $A$ to $B$ and the product is over all sites belonging to path $\mathcal P$. {For $W\gg t=1$}, the weight $\prod_{i\in\mathcal P} (E_F-\varepsilon_i)^{-1}$  of a path $\mathcal P$ will decrease exponentially with its length. The sum over paths will then be dominated by the forward-scattering paths which propagate from left to right \cite{nguyen1985tunnel, fritzsche1990hopping, medina1989interference, prior2005conductance, PhysRevLett.99.116602, prior2009conductance, PhysRevB.93.054201}.
One thus obtains a mapping to DP where the partition function $  Z = \sum_{\mathcal {DP}} \prod_{i\in\mathcal {DP}} e^{-\beta V_i} $ corresponds to the {locator} expansion with on-site disorder $V_i = \ln (E_F-\varepsilon_i)$ and inverse temperature $\beta=1$.

Using this analogy and the universality of the physics of DP \cite{halpin1995kinetic, gueudre2015ground}, one can {infer} a certain number of properties of conductance fluctuations in the localized regime in dimension two: The logarithm of the conductance
$\ln g$ {should be} analogous to minus the free-energy of the DP problem and {thus should
follow:} 
$
 \ln g=-2 L/\xi~+~\alpha\left(L/\xi\right)^\theta \chi \; ,
$
with $\xi$ the localization length, $\theta=1/3$ the universal value of DP in dimension $d=2$ and $\chi$ a random variable of order one with the Tracy-Widom distribution. 
{These fluctuation properties have been precisely validated by numerical simulations at large scales $L\gg\xi$ \cite{prior2005conductance, PhysRevLett.99.116602, prior2009conductance, PhysRevB.91.155413}.
However, they do not reflect all the properties that AL could inherit from DP.} 

\emph{Pinning and avalanches.---} 
I now address the {unforeseen} glassy properties of {quantum transport in the presence of AL.
According to the analogy with DP,} the electron flow {should be} very inhomogeneous, and {follow} directed paths which correspond to the dominant paths of the DP problem \cite{halpin1995kinetic}. 
{As first proposed in \cite{pichard1991quantum, markovs2010electron},} it is possible to visualize such paths by considering how the conductance of a sample is affected when one changes the on-site energy $ \varepsilon_i $ of site $i$ by its opposite value $ - \varepsilon_i $. Thus, as seen in Fig.~\ref{fig:AndApath}, by representing in a color plot $ \delta g(i)=\vert g (i) -g_0\vert / g_0 $ as a function of the site position $ i $, where $ g_0 $ is the conductance of the sample without perturbation, and $ g(i)$ the conductance of the sample locally perturbed in $ i $, the flow of electron is visualized {(see \cite{supmat} for more details)}. This method is similar to {the experimental local probe approach called} scanning gate microscopy \cite{topinka2001coherent, abbout2011thermal, gorini2013theory}. {Moreover, it corresponds in {DP to the response of the free energy to a local perturbation on a site. This gives exactly \cite{maes2017midpoint} the probability that DP passes on the site.}}

\begin{figure}
\includegraphics[width=0.48\linewidth]{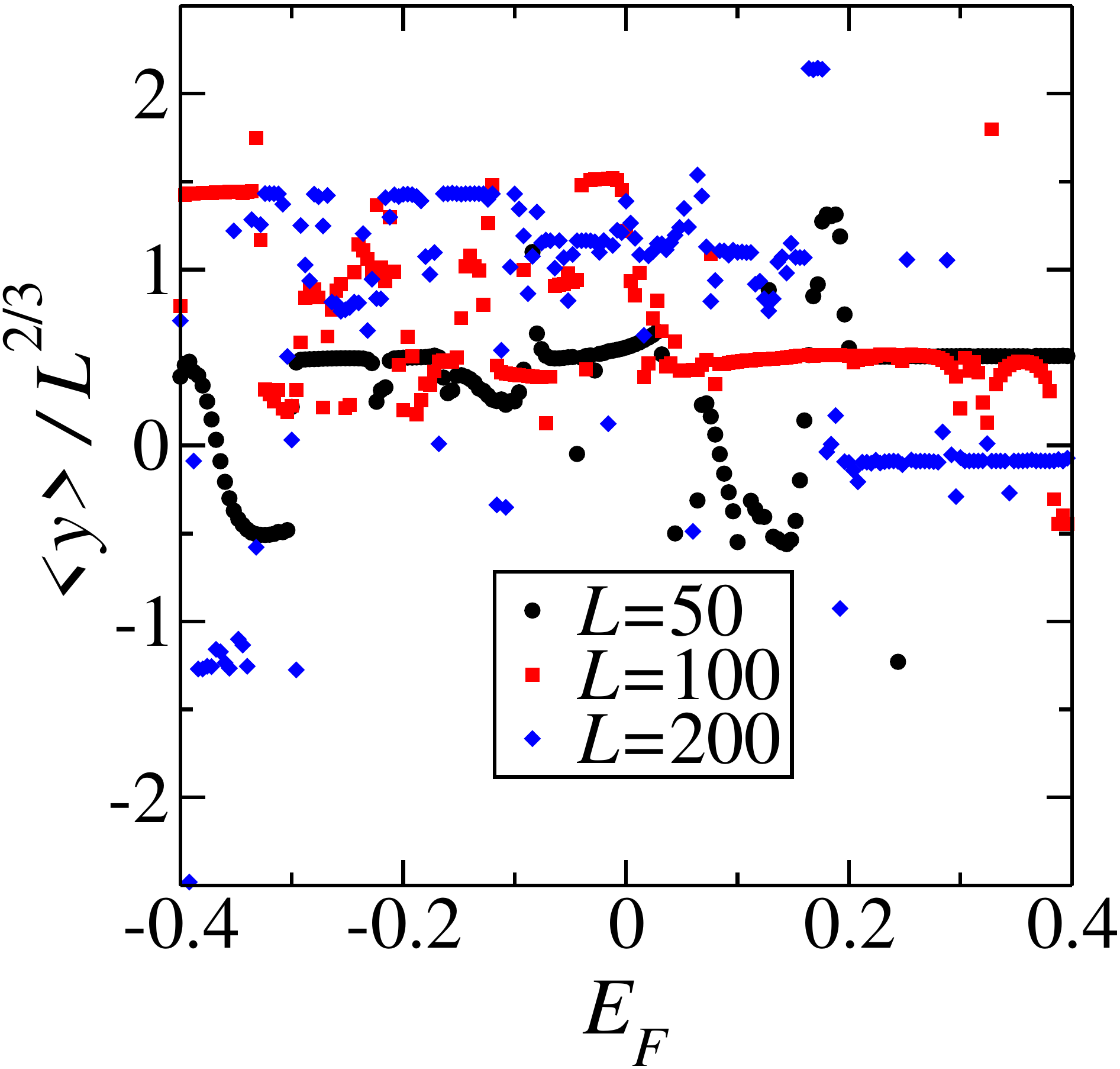}
\includegraphics[width=0.48\linewidth]{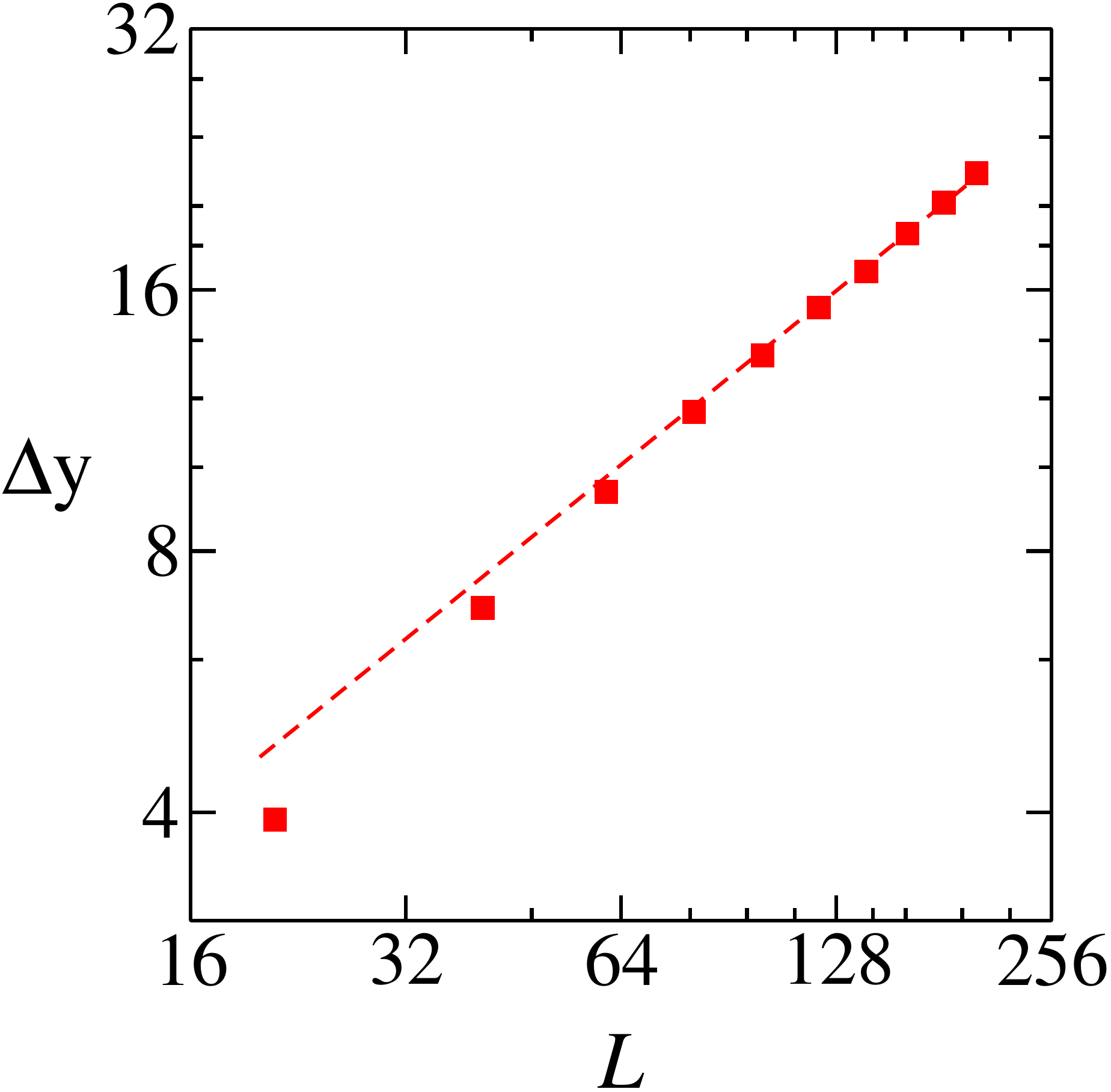}
 \caption{Pinning and avalanches of Anderson localization.
\emph{Left}: Similarly to DP, glassy properties arise from the sensitivity of the path taken by transport to a perturbation, such as a slight change of the  energy $E_F$. 
The final position of the path $\langle y \rangle$ (see text) is pinned most of the time but jumps abruptly to a very different value. These (equilibrium) avalanches make a brutal change between different paths such as those represented in Fig.~\ref{fig:AndApath}. The parameters are the same as in Fig.~\ref{fig:AndApath}. \emph{Right}: Pinning can be characterized by the roughness exponent $\zeta$ defined as $\Delta y = \sqrt{\overline{\langle y \rangle^2}} \sim L^{\zeta}$. For the DP problem $\zeta=2/3$ is exactly known. This behavior is represented by the red dashed line and agrees well with numerical data for the Anderson model with a normal distribution, $W=15$ {($\xi\approx 0.7$)}, and averaging over $18000 $ samples. The typical size of the jumps $\Delta y$ in the left panel is controlled by the roughness exponent.}
 \label{fig:AndAval}
\end{figure}

{Using this approach in the strongly localized regime $L\gg \xi$, I demonstrate the first two glassy properties already introduced:} pinning of the electron flow and avalanches between different directed paths when varying a parameter such as the  energy $ E_F $.
In Fig.~\ref{fig:AndAval}, I represent {as a function of $ E_F $, the final transverse position $\langle y \rangle$ of the path taken by transport}
\begin{equation}{
 \langle y \rangle \equiv \frac{1}{\sum_y \delta g(L,y)} \sum_y y \; \delta g(L,y)  }   \; ,
\end{equation}
{in three different strongly disordered samples of sizes $L=50$, $100$ and $200$, box-distributed with $W=20$, i.e. $\xi\approx 1.5 \ll L$}. One can clearly observe plateaus which are the signature of pinning, with brutal jumps at certain values of $E_F$ which depend on the microscopic disorder configuration. Thus, in the case of Fig.~\ref{fig:AndApath}, the path taken by the electron flow  remains the same as that of the left panel between $E_F = 0$ and $E_F = 0.1$ and suddenly jumps to the path of the right panel at $E_F = 0.1$ up to $E_F = 0.2$.

In the DP problem, pinning is characterized by the roughness exponent $\zeta$ which measures the wandering of the DP:
$\zeta$ is defined through $\overline{\langle y \rangle^2} \sim L^{2 \zeta}$ for a point-like initial condition starting at {$y=0$} for $x=0$ ($\overline{X}$ means disorder averaging of $X$). In dimension $d=2$, $\zeta=2/3$ is known exactly. The Fig.~\ref{fig:AndAval}, right panel, represents the     
evolution with system size of $\Delta y = \sqrt{\overline{\langle y \rangle^2}}$ for the 2D AL problem {in the strongly localized regime $L\gg \xi$}. The numerical data agree well with the DP value $\zeta=2/3${, confirming the conjecture of \cite{medina1989interference, PhysRevB.46.9984} based on the locator expansion}. Moreover, in the left panel of Fig.~\ref{fig:AndAval}, the typical size of the equilibrium avalanche jumps is seen to scale as $\Delta y \sim L^{2/3}$, as expected for DP \cite{mezard1990glassy}.
Together with $\theta=1/3$ \cite{prior2005conductance, PhysRevLett.99.116602, prior2009conductance}, $\zeta=2/3$ confirms that {these large scale properties} of 2D AL belong to the same universality class as that of the DP problem.

\begin{figure}
\includegraphics[width=\linewidth]{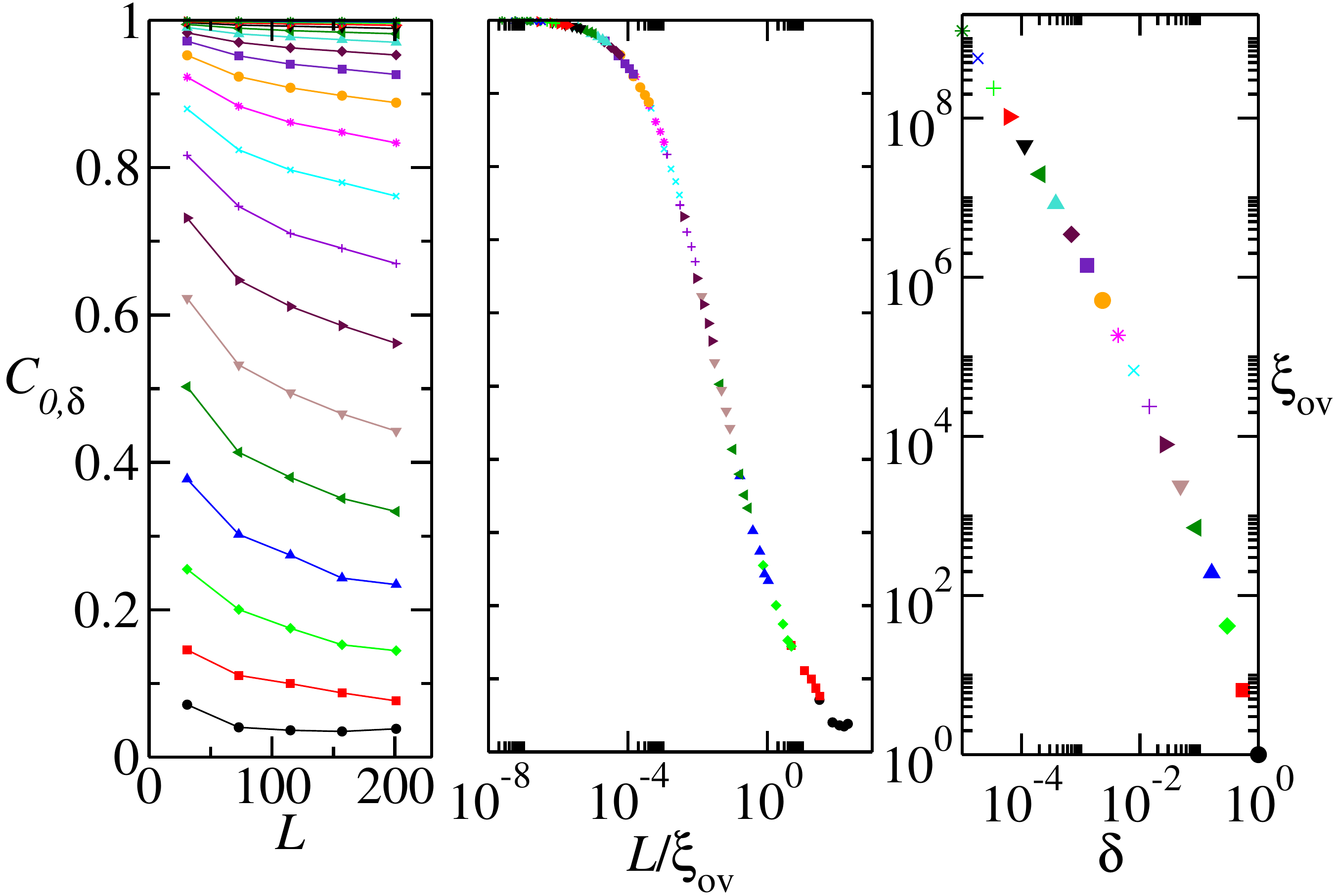}
 \caption{Disorder chaos of Anderson localization. The correlation function of $\ln g$ with a perturbation of the disorder configuration \eqref{eq:pertdis} characterized by the parameter $\delta$ is plotted as a function of system size $L$ in the left panel (decreasing $\delta$ from upper to lower curves). The middle panel shows the single parameter scaling of $\mathcal C_{0,\delta} (L)$ as a function of the overlap length $\xi_\text{ov}(\delta)$. This scaling function shows that an infinitesimal perturbation induces large changes in the conductance so that $\mathcal C_{0,\delta} (L)$ vanishes for $L\gg \xi_\text{ov}$. The divergence of $\xi_\text{ov}$ as a function of $\delta$ is shown in the right panel. $72 000$ disordered 2D samples connected to a narrow left lead and a wide right lead have been considered for each values of $\delta$ and $L$, with a disorder strength $W=10$ {($\xi \approx 1.$)} and $E_F=0.01$.}
 \label{fig:C-And-1DLlead}
\end{figure}

\emph{Chaos.---} I now turn to the spectacular glassy effect known as chaos \cite{mckay1982spin, PhysRevLett.58.57, PhysRevB.38.386, PhysRevB.43.10728, sales2002fragility, da2004temperature}. I show that two replicas of the Anderson model, infinitesimally perturbed with respect to each other, have completely uncorrelated conductances at the thermodynamic limit, i.e. {the conductance} is fragile.
In close analogy with the disorder chaos effect in spin glasses and DP, I study the change of the conductance under a perturbation of the disorder configuration:
\begin{equation}\label{eq:pertdis}
 \varepsilon^\delta_i = \frac{\varepsilon_i + \delta \varepsilon'_i}{\sqrt{1+\delta^2}}\; .
\end{equation}
Here, $\varepsilon$ and $\varepsilon'$ are both normally distributed with the same mean $0$ and standard deviation $W$ and $\delta$ denotes the strength of the perturbation. I consider the correlation function of the logarithm of the conductance between the two replicas:
\begin{equation}\label{eq:correl}
 \mathcal C_{0,\delta} (L) = \frac{\overline{\Delta \ln g_\delta \; \Delta \ln g_0}}{\sqrt{\overline{{\Delta \ln g_\delta}^2}}\sqrt{\overline{{\Delta \ln g_0}^2}}} \; ,
\end{equation}
where $\Delta \ln g \equiv \ln g - \overline{\ln g}$.

In the left pannel of Fig.~\ref{fig:C-And-1DLlead}, $\mathcal C_{0,\delta} (L)$ is represented as a function of the system size $L$ for different values of the perturbation strength $\delta$. In the limit of small $\delta$ (upper curves), the two replicas are strongly correlated, $C_{0,\delta}\approx 1$ {while} at large $\delta$, the correlation is vanishing. The middle panel shows that all the data collapse onto a single curve when plotted as a function of $L/\xi_\text{ov}$, with $\xi_\text{ov}$ known as the overlap length. 
This scaling behavior supports the chaos property of AL: {an infinitesimal perturbation induces a vanishing correlation for $L\gg \xi_\text{ov}$.}

\begin{figure}[t]
\includegraphics[width=\linewidth]{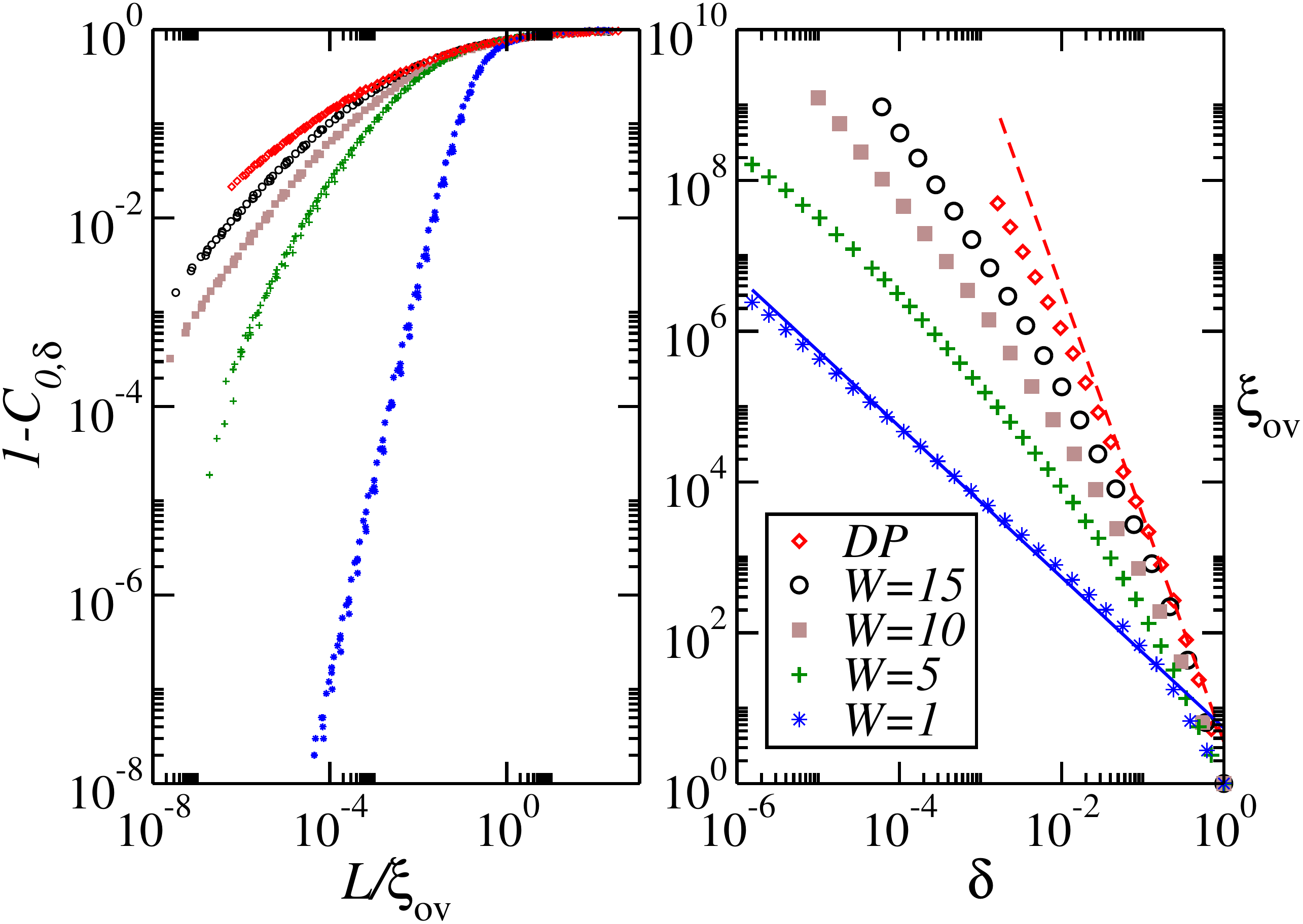}
 \caption{Disorder chaos of Anderson localization for different disorder strengths $W$. The scaling function and the overlap length divergence depend crucially on {$W$, contrary to DP}. The red dashed line corresponds to the droplet scaling {argument} $\xi_\text{ov} \sim \delta^{-3}$, which agrees well with the DP data and the AL data at large $W$ and not too small $\delta$.
 The blue line indicates the expected behavior in the weak localization regime $\xi_\text{ov} \sim \delta^{-1}$ (see text), which agrees well with the data at $W=1$ for small $\delta$. Otherwise, identical to Fig.~\ref{fig:C-And-1DLlead}}
 \label{fig:C-And-1DLlead-vsW}
\end{figure}

The overlap length $\xi_\text{ov}$ depends only on $\delta$ and is shown in the right panel of Fig.~\ref{fig:C-And-1DLlead}.
{In the DP problem, its} divergence can be understood from a droplet scaling argument \cite{PhysRevB.43.10728, sales2002fragility}. 
{To this end, I consider the case of a DP with real on-site energies $V_i = \ln\vert \varepsilon_i\vert$ (such an argument is not available for complex $V_i = \ln (E_F- \varepsilon_i)$)} with $E_F=0$. The scaling argument compares the energy cost of remaining in the same path in the perturbed replica with the free-energy gain to jump into another configuration $\Delta F \sim L^{1/3}$. Here, the energy cost can be written as:
\begin{equation}
 \Delta E = \sum_{i=1}^{L} \ln(\vert \varepsilon^\delta_i \vert ) - \ln(\vert \varepsilon_i \vert) = \sum_{i=1}^{L} \ln \frac{\left\vert 1+\delta \frac{\varepsilon'_i}{\varepsilon_i} \right\vert}{ \sqrt{1+\delta^2}}\;,
\end{equation}
where the sum is along the sites of the dominant path. While $\varepsilon_i$ are correlated random variables (the dominant path is the result of a global optimization), the perturbed $\varepsilon'_i$ are uncorrelated. {We can thus neglect the correlations in the $\log$. Then, a} simple calculation shows that the variable $\ln \vert 1+\delta \varepsilon'_i/\varepsilon_i\vert$ has a finite average value $\ln \sqrt{1+\delta^2}$ and a standard deviation  $\sigma \propto \sqrt{\delta}$. According to the central limit theorem: $\Delta E \sim \sqrt{\delta L}$. Because $\Delta E $ and $\Delta F$ do not vary with the same power laws with $L$, {for $L\ll \xi_\text{ov}$, $\Delta E \ll \Delta F$} and the two replicas are in the same configurations, whereas {for $L\gg \xi_\text{ov}$, $\Delta E \gg \Delta F$}, the perturbation has made the path jump into a completely different configuration and the replicas are uncorrelated. 
The overlap length is such that $\Delta E \sim \Delta F$, i.e. $\delta^{1/2} {\xi_\text{ov}}^{1/2} \sim {\xi_\text{ov}}^{1/3}$, therefore $\xi_\text{ov} \sim \delta^{-3}$.
In Fig.~\ref{fig:C-And-1DLlead-vsW}, I represent the disorder chaos effect on such DP with $V_i = \ln\vert \varepsilon_i\vert$ (diamond points). There is a good agreement between the data for $\xi_\text{ov}$ and the prediction of the droplet scaling argument {(red dashed line)}. 

{While AL data are quite close to the DP results at the strongest disorder $W=15$ I could consider (open circles in Fig.~\ref{fig:C-And-1DLlead-vsW}), a} striking observation of Fig.~\ref{fig:C-And-1DLlead-vsW} is however that disorder chaos for AL depends crucially on $W$, with a stronger fragility as $W$ decreases from $15$ to $1$. The strong fragility at small disorder $W=1$ corresponds to the famous universal conductance fluctuations \cite{Akkermans:book07, PhysRevB.35.1039, altland1994electronic}. {Then,} the conductance results from the interference between many multiply scattered Feynman paths {(instead of one main path as before)} and thus crucially depends on their relative phases. These phases can be altered by a perturbation. In the case of a change of $E_F$ {by $\omega$}, 
a pair of identical Feynman paths dephases by an amount $\omega t_{\mathcal P}/\hbar$ where $t_{\mathcal P}$ is the time to {travel} the path. In the diffusive regime, $t_{\mathcal P}\propto {E_\text{Th}}^{-1}$ with $E_\text{Th} = \hbar D/L^2$ the Thouless Energy, $D$ the diffusion constant.
Therefore, the overlap length follows {$\xi_\text{ov} \sim {\omega}^{-1/2}$} {(see \cite{supmat})}.
In the {present} case of a disorder perturbation \eqref{eq:pertdis}, the dephasing is proportional to $\delta \sqrt{t_{\mathcal P}}$ due to \emph{random} shifts of alternative signs {(instead of the constant shift $\omega$)} of the on-site energies. This explains the observed behavior $\xi_\text{ov} \sim \delta^{-1}$, shown by the blue line in the right panel of Fig.~\ref{fig:C-And-1DLlead-vsW}.

The observations of Fig.~\ref{fig:C-And-1DLlead-vsW} are {therefore} quite different from DP for which the chaos effect does not depend on the strength of disorder \cite{sales2002fragility}.
Interestingly, the chaos property of AL results from the interplay between two distinct mechanisms: an intereference effect {between many paths} which {dominates} in the weak localization regime \cite{Akkermans:book07},  and the glassy chaos effect of DP {at strong disorder which brutally changes the directed path taken by the transport. {This persists at large scales $L\gg \xi$, since the scaling function itself depends on $W$. Chaos gives a new light on the analogy between AL and DP: the directed paths taken by transport are a coarse-grained picture. They have a width of order $\xi$ and quantum interference play a crucial role inside these paths \cite{supmat}.}

\emph{Experimental observations.---}
The glassy effects described here could be observed experimentally in several types of systems such as two-dimensional electron gases \cite{abrahams201050, Bergmann:PR84}, cold atom systems, or classical waves. e.g. microwaves \cite{genack2010speckle} or ultrasounds \cite{lagendijk2009fifty}.
In two-dimensional electron gases, {scanning gate microscopy \cite{topinka2001coherent} could image the directed paths associated with AL}. In cold atomic gases, AL {has been} characterized in a very controlled manner \cite{sanchez2010disordered,aspect2009anderson, lopez2013phase, PhysRevLett.115.240603}. Very recently, an experimental approach to study the conductance in cold atoms has been developed \cite{0953-8984-29-34-343003}, which allows the introduction of disorder \cite{PhysRevLett.115.045302} and scanning gate microscopy imaging \cite{PhysRevLett.119.030403}. 

\emph{Conclusion.---}
In this article, I have shown from extensive numerical simulations that AL in two dimensions and at zero temperature satisfies several important glassy properties {inherited from DP physics but where quantum interference effects play a crucial role: pinning, avalanches, and chaos. AL opens a new playground for the study of quantum glassy physics.}  
 
It would be particularly interesting to see whether these glassy properties extend to the case of interacting systems. The method of analysis presented here could in particular help to clarify the glassy nature of the Bose glass insulating phase of disordered bosons \cite{Fisher:dirtyboson:PRB89, Giamarchi1988a, Altman2008a, doggen2017weak, muller2013magnetoresistance, gangopadhyay2013magnetoresistance, monthus2012random}.

\begin{acknowledgments}
I thank the late Jean-Louis Pichard for many inspiring discussion.
I thank also M. Dupont, F. Evers,
N. Laflorencie, C. Monthus, and A. Rosso for interesting
discussions.
I thank CalMiP for access to its supercomputer.
This work was supported by Programme Investissements
d'Avenir under the program ANR-11-IDEX-0002-02, reference  ANR-10-LABX-0037-NEXT, and the ANR grant COCOA No ANR-17-CE30-0024-01.
\end{acknowledgments}


%\newpage

%******************************************************************************
%******************************************************************************
%\section{Supplemental material}
%******************************************************************************
%******************************************************************************

\onecolumngrid
\newpage
\appendix
\begin{center}
{\large\textbf{Supplemental material to\\ ``Glassy properties of Anderson localization: pinning, avalanches and chaos.''
}}
\end{center}
%%%%%%%%%% Prefix a "S" to all equations, figures, tables and reset the counter %%%%%%%%%%
\setcounter{equation}{0}
\setcounter{figure}{0}
\setcounter{table}{0}
\setcounter{page}{1}
\makeatletter
\renewcommand{\theequation}{S\arabic{equation}}
\renewcommand{\thefigure}{S\arabic{figure}}
\renewcommand{\bibnumfmt}[1]{[S#1]}
\renewcommand{\citenumfont}[1]{S#1}

\twocolumngrid

\section{Scattering configurations}
\begin{figure}[b]
\includegraphics[width=0.7\linewidth]{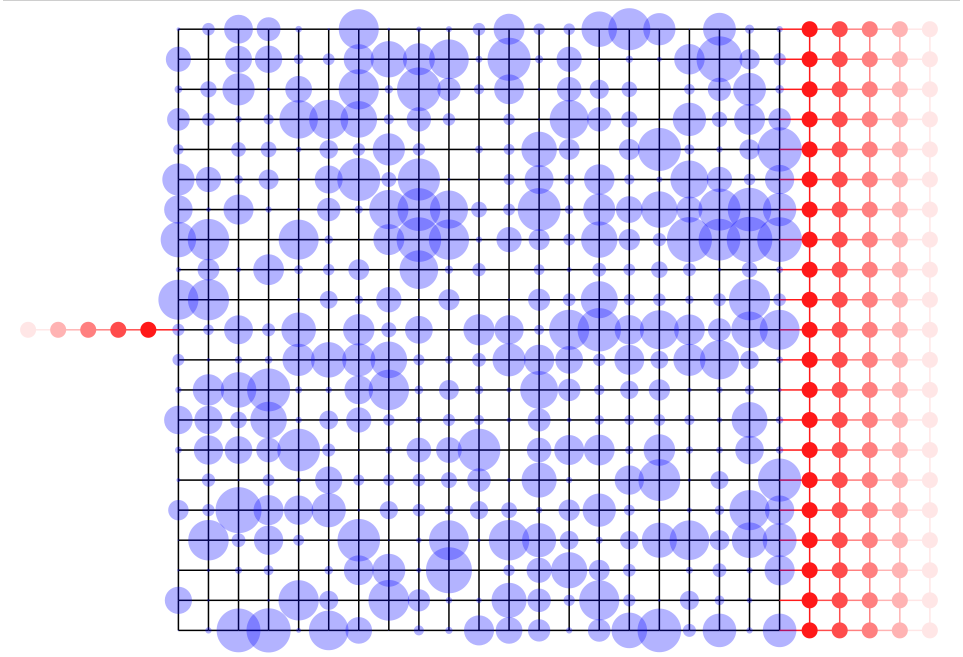}
\includegraphics[width=0.7\linewidth]{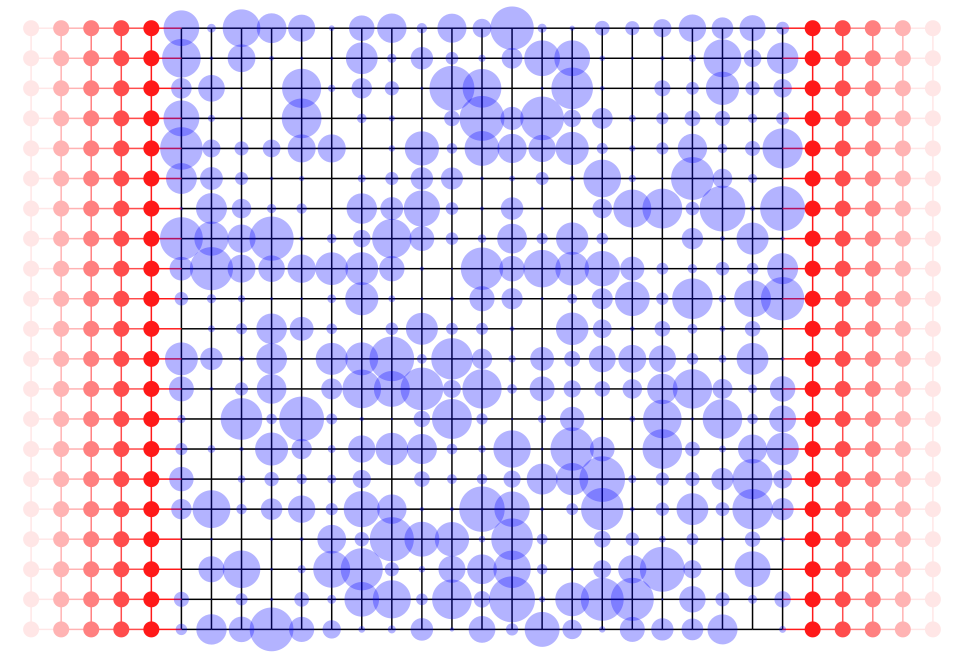}
\caption{Illustration of the two scatterering configurations considered: the upper panel corresponds to the narrow left lead configuration, where a 1D semi-infinite lattice is attached at the middle (defined as $y=0$) of the left edge (with $y \in [-L/2,L/2]$) of the scattering region. The lower panel shows the wide left lead case, where the width of the left lead is the same as that of the scatterer. The quantum disordered system is shown in violet, with point sizes following the value of the on-site energy $\varepsilon_i$ (see Eq.~(1)). The leads are shown in red.}
 \label{fig:1DWLlead}
\end{figure} 

Two different scattering configurations have been considered. A quantum disordered system described by the 2D Anderson tight-binding model (1) (shown in violet in figure \ref{fig:1DWLlead}) is attached to two perfect, semi-infinite (along the $x$-direction) leads which can be wide, having the same width $L$ of the scatterer, or narrow, consisting in a 1D semi-infinite lattice attached at the middle ($y=0$) of one edge of the system (with $y \in [-L/2,L/2]$). The figure \ref{fig:1DWLlead} illustrates these two scattering configurations: the disordered scattering region is represented in violet with the size of the points chosen according to the value of the on-site energy $\varepsilon_i$, and the leads are shown in red. All the results presented in the article have been obtained using the first narrow left lead configuration, while similar results in the case of a wide left lead will be shown in this supplemental material.

\section{Conductance fluctuations in the strongly localized regime}

 \begin{figure}
\includegraphics[width=\linewidth]{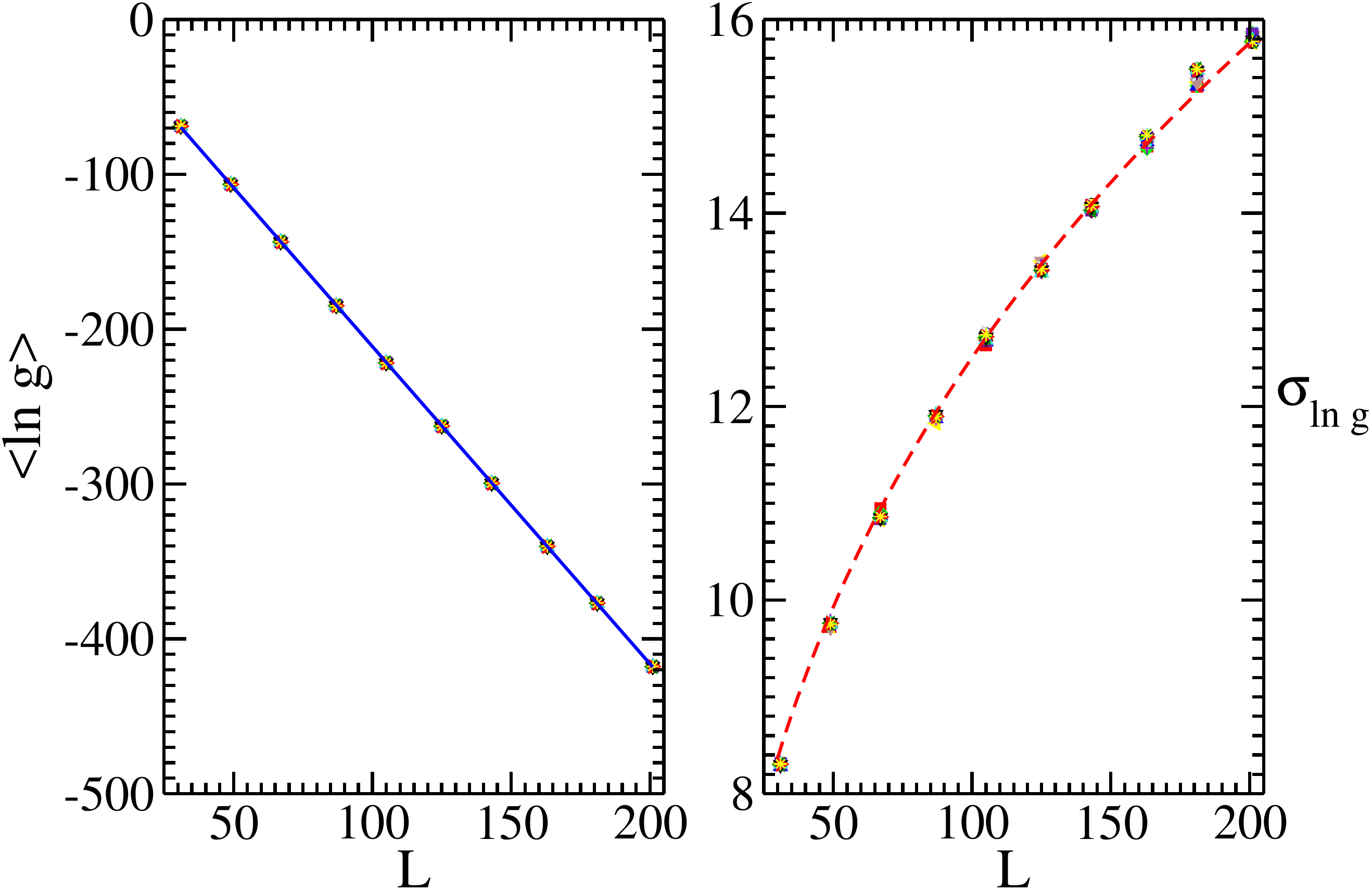}
\includegraphics[width=0.9\linewidth]{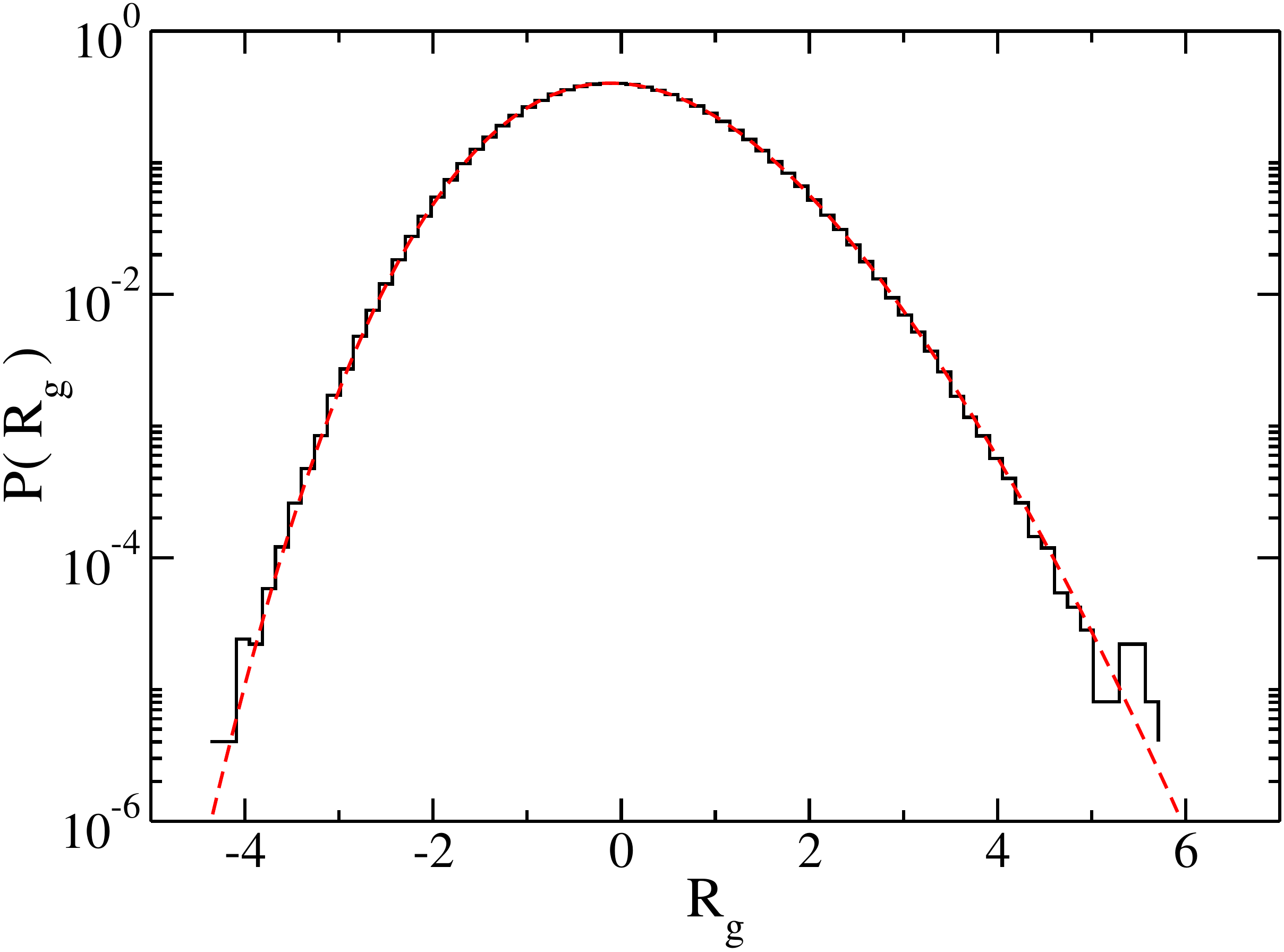}
\caption{Conductance fluctuations in the strongly localized regime. The left upper panel represents $\overline{\ln g}$ as a function of system size $L$, where the blue line is a linear fit $\overline{\ln g} = A0+2 L/\xi$ with a fitted localization length $\xi \approx 1$. The right upper panel shows the standard deviation $\sigma_{\ln g} = \sqrt{\overline{(\ln g - \overline{\ln g})^2}}$ as a function of the system size $L$. A very good agreement is found with the expected behavior  $\sigma_{\ln g}\sim L^{1/3}$ from the analogy with the directed polymer problem (red dashed line).
The lower panel shows the distribution of the rescaled conductance $R_g = (\ln g - \overline{\ln g})/\sigma_{\ln g}$ . A perfect agreement is found with the GUE Tracy-Widom distribution expected for the directed polymer problem with a point like initial condition. In the two upper panels, different points correspond to different values of the energy $E_F \in [0,0.4]$, and the data have been averaged over $72.10^3$ configurations, with a normally distributed disorder and $W=10$. In the lower panel, $E_F=0.01$, $W=10$, $L=100$ and $36.10^4$ disorder configurations have been considered. In all cases presented, a scattering configuration with a narrow left lead has been considered.}
 \label{fig:condfluctu}
\end{figure} 

\begin{figure*}
\includegraphics[width=0.7\linewidth]{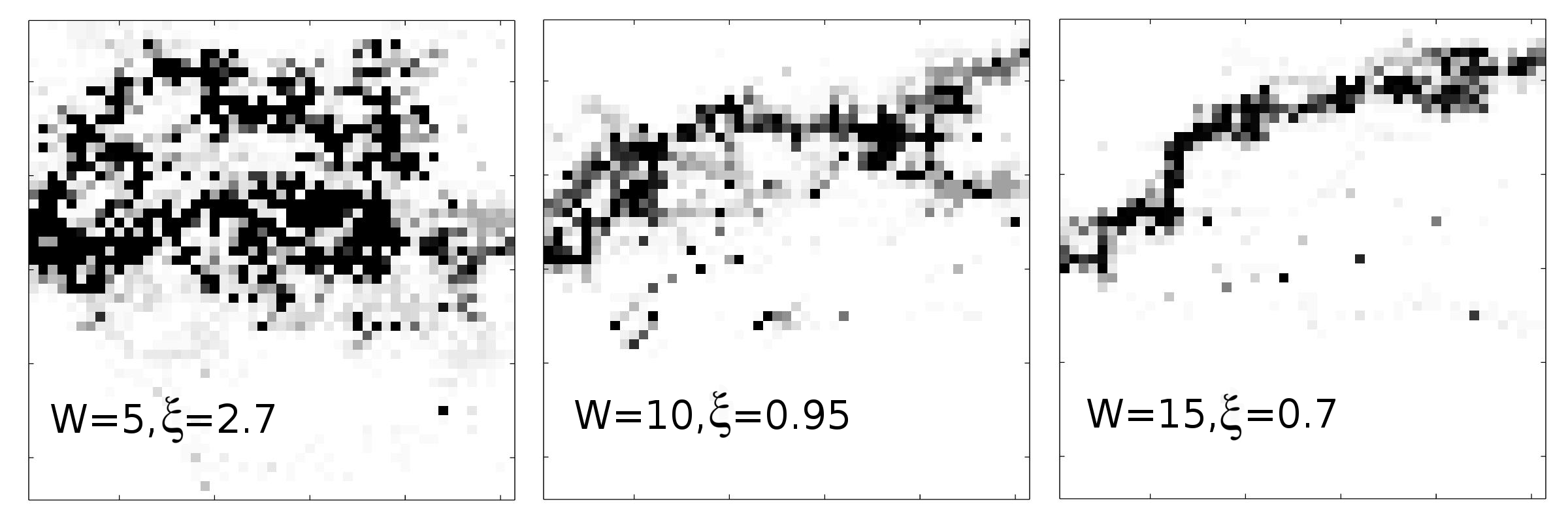}
\caption{Dependence on the strength of the disorder of the width of the directed paths taken by the transport. The directed paths taken by the transport have a width of order the localization length $\xi$. Inside these paths, quantum interference effects play a crucial role. $E_F=0$, $L=51$ and narrow left lead has been considered. }
 \label{fig:pathsWidth}
\end{figure*}

In the strongly localized regime, Somoza, Ortu\~no, and Prior \cite{PhysRevLett.99.116602, prior2005conductance, prior2009conductance} showed numerically that the conductance $g$ follows:
 \begin{equation}\label{eq:condfluctu}
  \ln g=-2 L/\xi~+~\alpha\left(L/\xi\right)^\theta \chi \; .
 \end{equation}
 In this equation, the first term describes the exponential decay of the typical conductance $\exp (\overline{\ln g} )= \exp (-2L/\xi)$ with $\xi$ the localization length.
 The second term of \eqref{eq:condfluctu} describes the fluctuations of $\ln g$ with $\chi$ a random variable of order $1$, $\alpha$ a constant of order $1$. The value of the exponent $\theta$ has been found to perfectly agree with the directed polymer universal value: $\omega=1/3$ in dimension $1+1$ \cite{halpin1995kinetic}. Moreover, the variable $\chi$ follows the Tracy-Widom distribution for the Gaussian unitary ensemble (GUE) in the case of a narrow left lead, which is again the known distribution for the directed polymer problem in dimension $1+1$ with a point-like initial condition \cite{johansson2000shape, tracy1994level, tracy1996orthogonal}. 
 
 The figure \ref{fig:condfluctu} illustrates these known results. I represent the typical conductance and the standard deviation of $\ln g$ as a function of the system size $L$ in the case of a narrow left lead and a normally distributed disorder with $W=10$. A very good agreement is found with \eqref{eq:condfluctu}, with a fitted localization length $\xi\approx 1$ and $\omega=1/3$. Moreover, the rescaled variable $R_g = (\ln g - \overline{\ln g})/\sigma_{\ln g}$ follows the GUE Tracy-Widom distribution shown by the red-dashed line (with no fitting parameter). 
 
 Note that in Fig. \ref{fig:condfluctu}, different symbols correspond to different values of the energy $E_F \in [0,0.4]$. The fact that they all have the same behavior shows that the glassy properties observed with a change of $E_F$ between $E_F=0$ to $0.4$ are not due to a significant change of the localization length. On the contrary, these changes are related to the large fluctuations induced by disorder and quantum interference, which grow with system size.

 An important remark is that the results of Somoza, Ortu\~no, Prior and Le Doussal \cite{PhysRevLett.99.116602, prior2005conductance, prior2009conductance, PhysRevB.91.155413} could suggest that the analogy between Anderson localization and the physics of directed polymer becomes asymptotically exact in the thermodynamic limit $L \gg \xi$. This is at first surprising since the directed polymer problem is a fully classical problem.  
 What I have shown is that quantum interference effects play a crucial role in the glassy properties of Anderson localization, even in the limit $L\gg \xi$. This can be seen from the chaos property. The fact that the scaling functions in Fig. 4 depend crucially on $W$, going from weak localization fragility at weak disorder to spin glass chaos at strong disorder, implies that these differences will survive in the thermodynamic limit $L \gg \xi$. My interpretation is that the directed paths taken by the transport (which follow the directed polymer physics) are a coarse-grained picture. They have a width of order the localization length $\xi$ and quantum interference effects play a crucial role inside these paths (see figure \ref{fig:pathsWidth}). 
 
 These results are not in contradiction with the known results for the fluctuations of the conductance in dimension two \cite{PhysRevLett.99.116602, prior2005conductance, prior2009conductance, PhysRevB.91.155413}. 
 The universal conductance fluctuations in the weak localization regime predict $\sigma_{\ln g}$ to be of order 1, whereas the directed polymer physics predicts much stronger fluctuations $\sigma_{\ln g} \sim L^{1/3}$. Directed polymer physics thus completely dominates the fluctuation properties of the conductance in the large scale regime. On the contrary, weak localization fragility is much stronger than spin glass chaos so that the importance of quantum interference effects is clearly seen using this glassy point of view. One can therefore see Anderson localization as a quantum glassy system, with new glassy properties dressed by quantum interference to be explored.

\section{Chaos with a change of $E_F$. Case of a wide left lead scattering configuration}

I represent here additional numerical results for the chaos property of Anderson localization in the cases of a change of $E_F$ or with a wide left lead (see figure \ref{fig:1DWLlead}).
In the figure \ref{fig:chaosEf} I consider the correlation of the logarithm of the conductance $C_\omega$ between two replicas (having the same disorder configuration) taken at two different energies, $E_f=0$ for replica A and $E_F=\omega$ for replica $B$:
\begin{equation}
 C_\omega (L) = \frac{\overline{\Delta \ln g_\omega \; \Delta \ln g_0}}{\sqrt{\overline{{\Delta \ln g_\omega}^2}}\sqrt{\overline{{\Delta \ln g_0}^2}}} \; ,
\end{equation}
where $g_\omega = g(E_F = \omega)$. $C_\omega$ decreases as a function of system size and follows a single parameter scaling law with the overlap length $\xi_\text{ov}$.
Similarly to the case of disorder chaos, the scaling function and parameter depend crucially on the disorder strength. For weak disorder, one recovers the expected behavior for the weak localization regime: $\xi_\text{ov}\sim \omega^{-1/2}$. In the limit of strong disorder, the behavior is found close to that of the directed polymer problem.

\begin{figure}
\includegraphics[width=\linewidth]{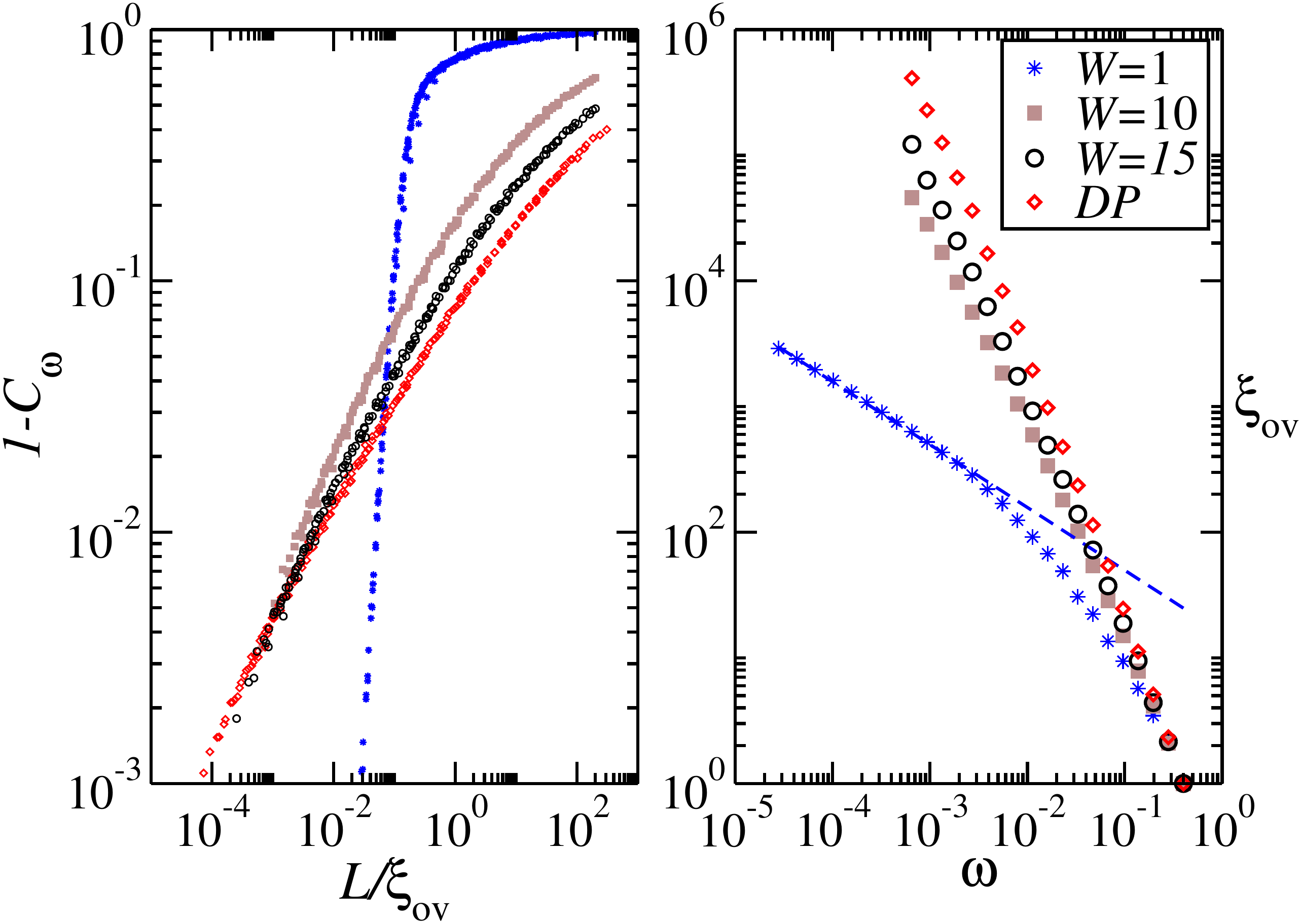}
\caption{Chaos property of Anderson localization with a change of the energy $E_F$. The correlation function $C_\omega$ is plotted
as a function of system size $L$ in the left panel. It follows a single parameter scaling as a function of the overlap
length $\xi_\text{ov}(\omega)$. The divergence of $\xi_\text{ov}$ as
a function of $\omega$ is shown in the right panel. As in the case of disorder chaos (figure 4 of the main paper), the scaling function and parameter both depend crucially on the disorder strength.
At weak disorder $W=1$, $\xi_\text{ov}\sim \omega^{-1/2}$ as expected in the weak localization regime (blue dashed line). At strong disorder, the behavior tends to the directed polymer problem.
$E_F=0$, $72.10^3$ disorder configurations connected to a narrow left lead have been considered. }
 \label{fig:chaosEf}
\end{figure}

In the figure \ref{fig:chaosWLlead}, I represent the disorder chaos property of Anderson localization in the case of a wide left lead (see figure \ref{fig:1DWLlead}). Qualitatively similar results are observed as in the case of a narrow left lead.

 \begin{figure}
\includegraphics[width=\linewidth]{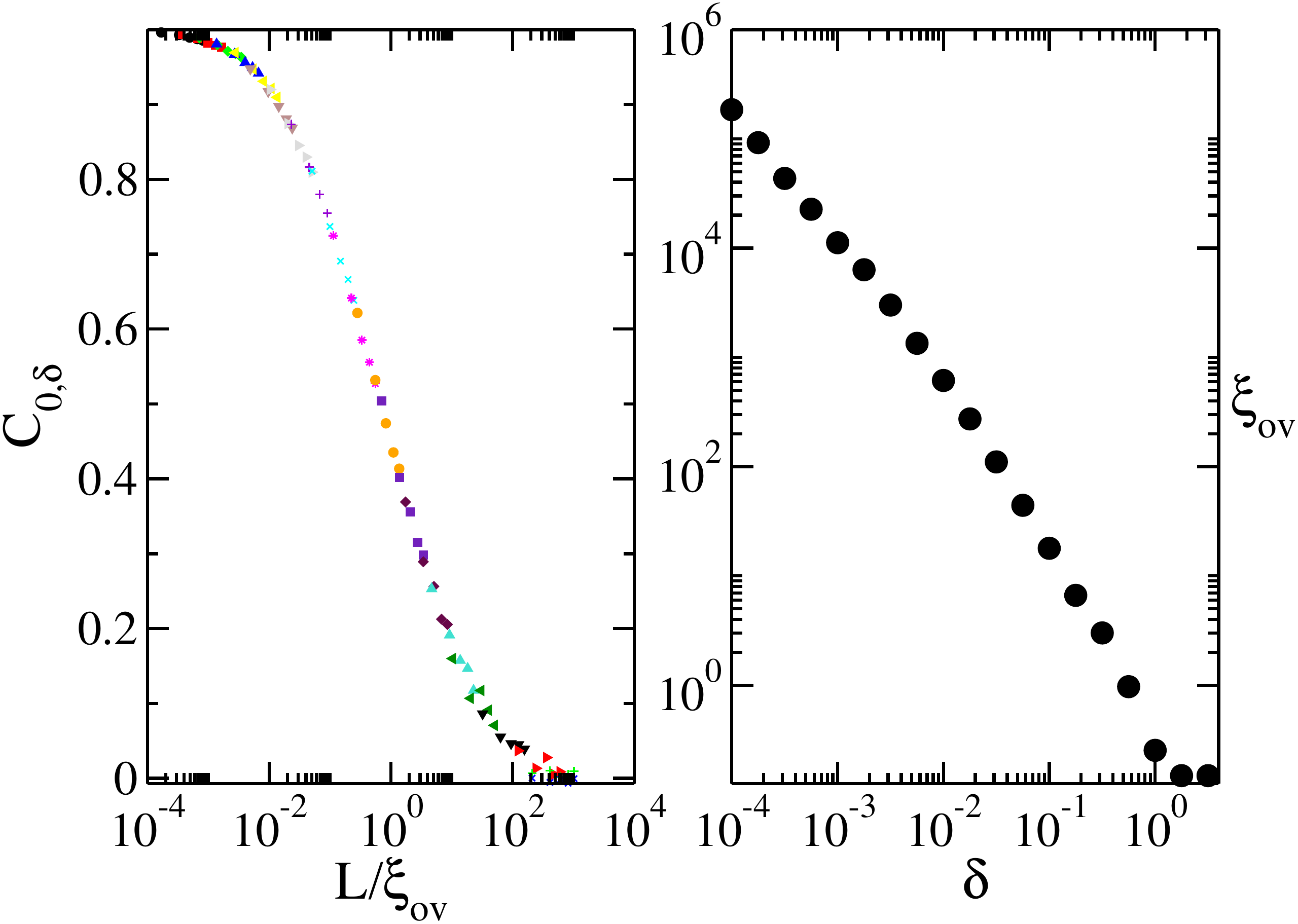}
\caption{Disorder chaos property of Anderson localization in the case of a wide left lead scattering configuration. Qualitatively similar results are observed as in the case of a narrow left lead, with single parameter scaling of the decay of the correlation $C_{0,\delta}$ with the scaling parameter $\xi_\text{ov}$ (left panel). The divergence of the overlap length $\xi_\text{ov}(\delta)$ at small $\delta$ is shown in the right panel. $E_F=0.01$, $W=10$ with a normal distribution, and $72.10^3$ samples have been considered.}
 \label{fig:chaosWLlead}
\end{figure}

\section{Scanning gate microscopy approach vs. local current density}

\begin{figure*}
\includegraphics[width=0.4\linewidth]{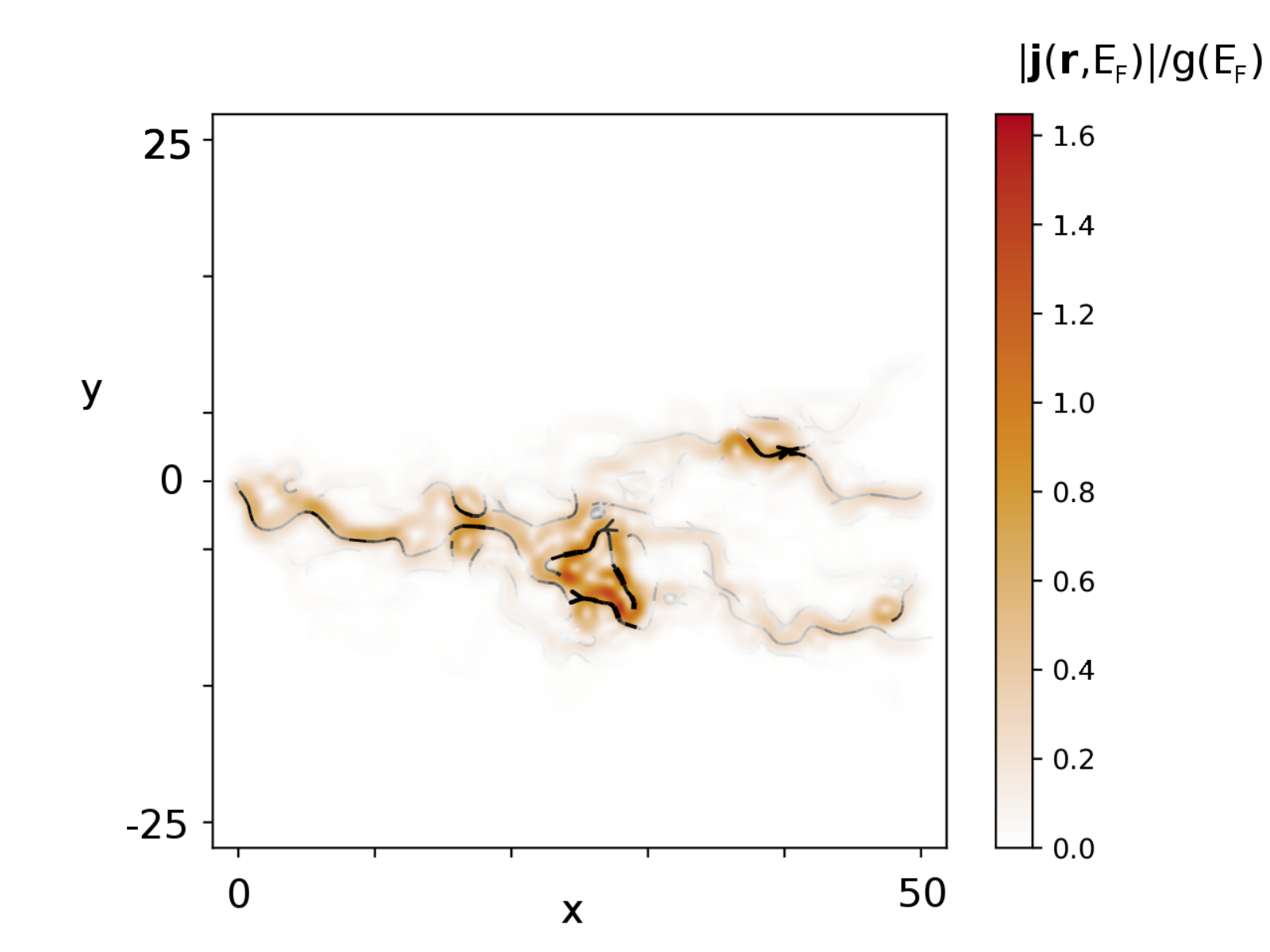}
\includegraphics[width=0.4\linewidth]{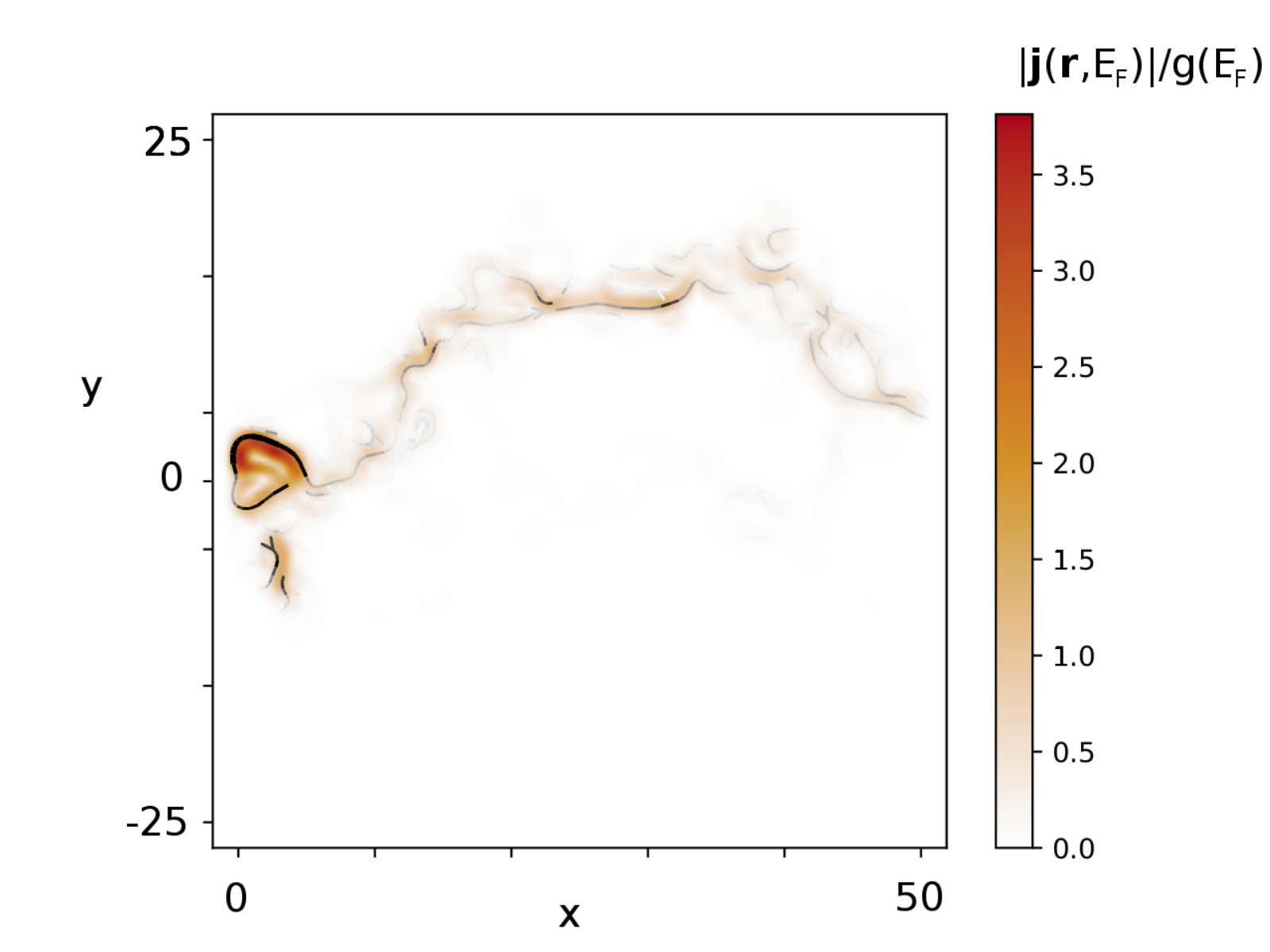}\\
\includegraphics[width=0.4\linewidth]{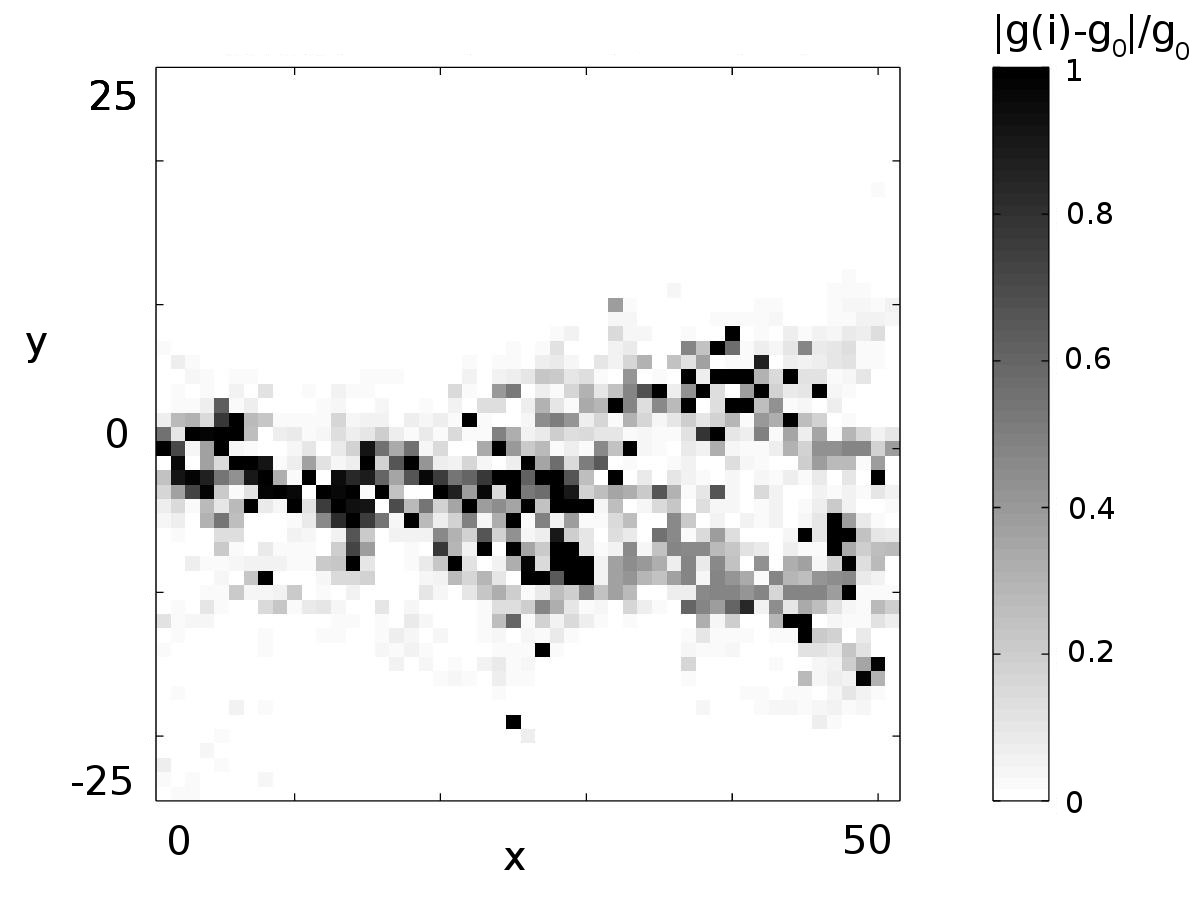}
\includegraphics[width=0.4\linewidth]{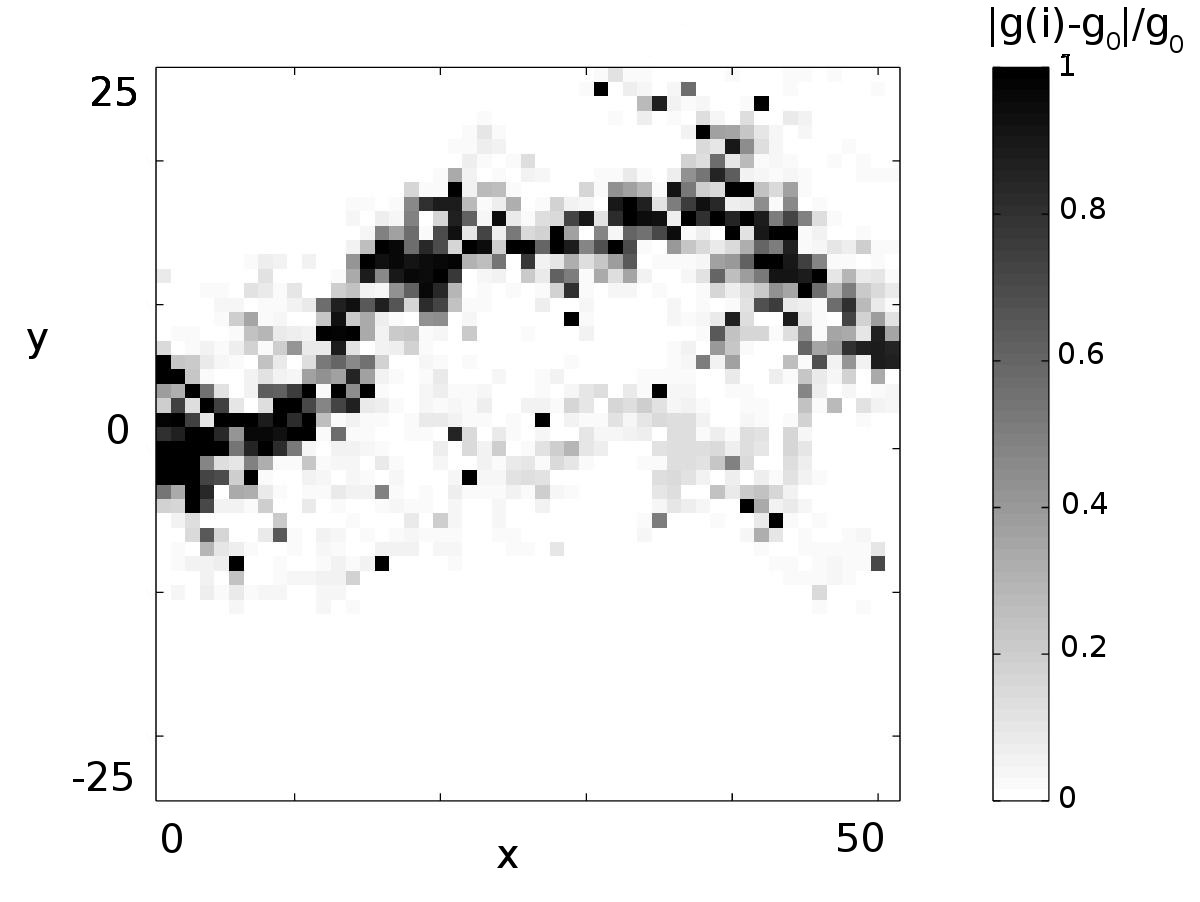}\\
\includegraphics[width=0.4\linewidth]{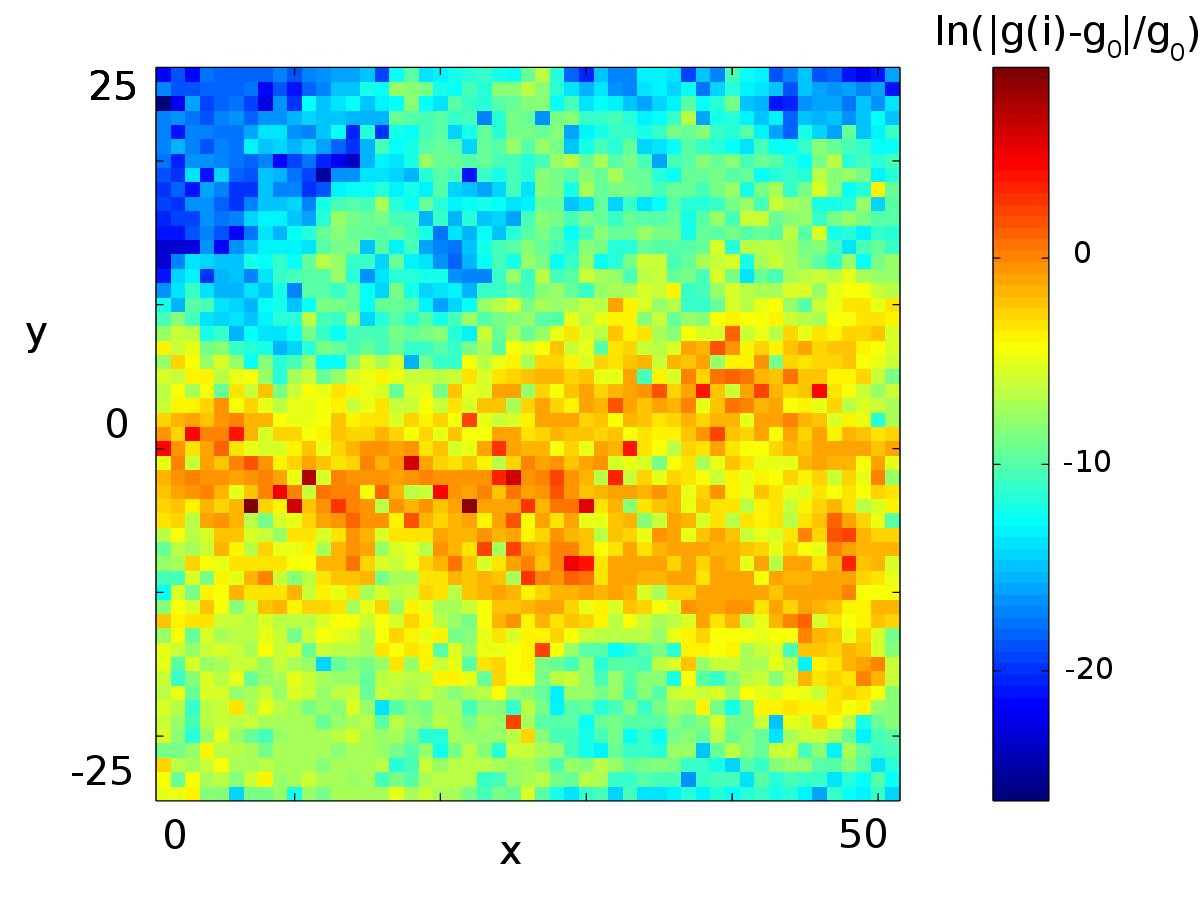}
\includegraphics[width=0.4\linewidth]{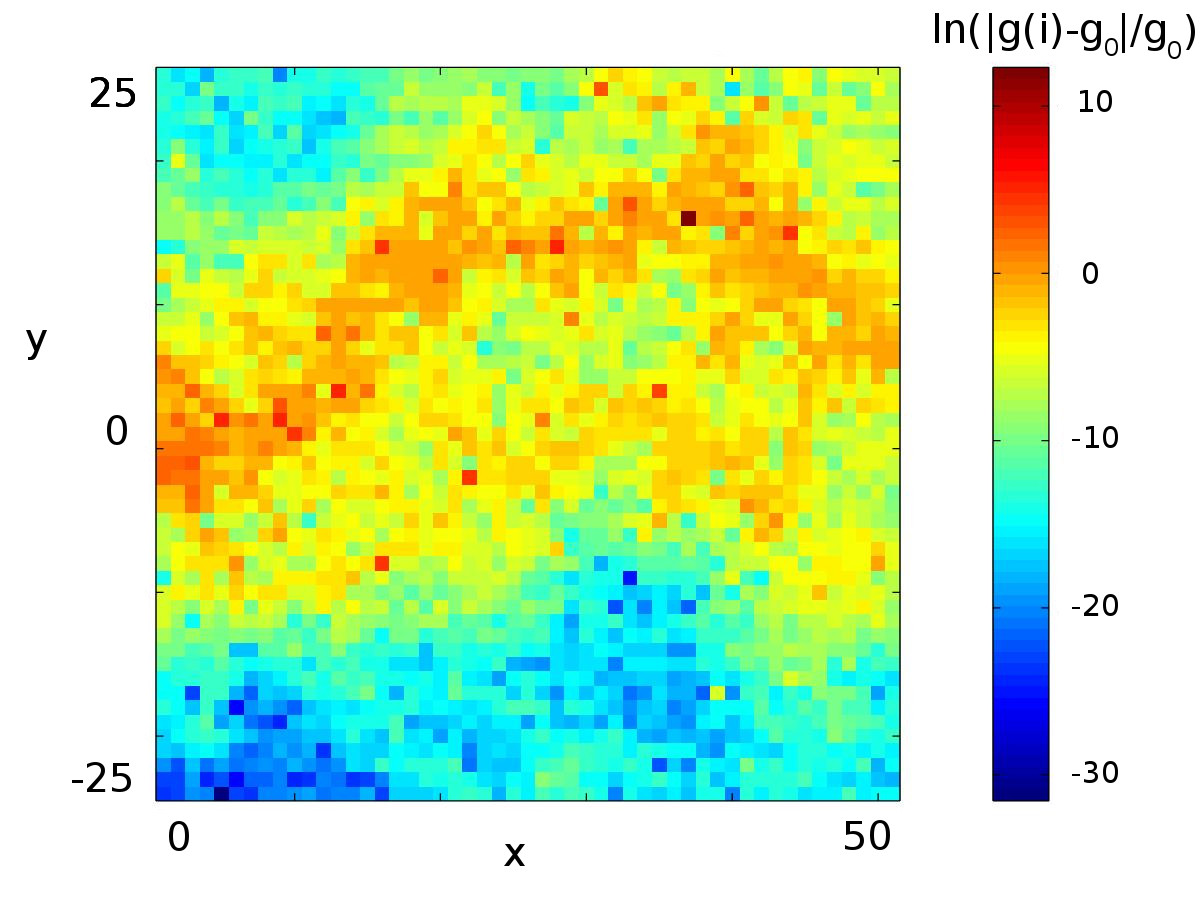}\\
\caption{Comparison between the local current density and the response of the conductance to a local perturbation.  The two upper panels show a map of the local current density for two different disordered samples. Below are shown the corresponding maps of the response of the conductance $\vert g(i) - g_0\vert/g_0$ to a local change of the on-site potential $\varepsilon_i \rightarrow -\varepsilon_i$, as used in the paper. In the middle panels, $\vert g(i) - g_0\vert/g_0$ is restricted to $[0,1]$ (see text) while a logarithmic scale with no restriction is shown in the lower panels. The paths imaged by the local conductance response clearly match with those of the local current density maps. This confirms the interpretation given in the paper that this method allows to image the paths taken by the electron flow. Samples of size $L=51$, with a box-distributed disorder and $W=10$, connected to a narrow left lead have been considered. The two upper panels have been obtained using the Kwant library \cite{groth2014kwant}.}
 \label{fig:currvsSGM}
\end{figure*}

In the paper, I have considered the response of $\ln g$ to a local change of the on-site potential $\varepsilon_i \rightarrow - \varepsilon_i$, as first proposed in \cite{markovs2010electron}. Then, by plotting $\vert g(i) - g_0\vert/g_0$ as a function of $i$, I get a color plot which indicates in dark the regions in space which are strongly affected by the local perturbation ($\vert g(i) - g_0\vert/g_0 \approx 1$), and in white the regions where this local change has a negligible effect $\vert g(i) - g_0\vert/g_0 \ll 1$. From the analogy between this approach and the scanning gate microscopy (SGM) experimental approach, on the one hand, and the correspondence with the exact result for the directed polymer probability to pass through a site \cite{maes2017midpoint}, I interpreted the paths obtained as the paths taken by the electron flow.

However, in the absence of disorder, it is known that SGM images are not always related to the local density of currents of the unperturbed system \cite{gorini2013theory}. Indeed, for a quantum point contact (QPC), SGM gives the local density of currents only on the plateaus of the conductance quantization where the quantum point contact is fully transparent. Otherwise, on the conductance steps, the SGM tip forms with the QPC a Fabry-Perot interferometer and the SGM images are not related to a local quantity of the unperturbed system such as the local current density \cite{gorini2013theory, abbout2011thermal}.

To clarify this point, I compare the map of the local current density $\textbf{j}(\boldsymbol{r}=(x,y), E_F)$ \cite{wilhelm2015ab}, obtained through the Kwant library \cite{groth2014kwant}, with the color plot given by the SGM approach I have used in the paper. In figure \ref{fig:currvsSGM}, I represent the case of two different disorder configurations, with a box-distribution and $W=10$. On rare points, the change $\vert g(i) - g_0\vert/g_0$ can be very large so that I restrict the colormap to $\vert g(i) - g_0\vert/g_0 \in [0,1]$ for the sake of clarity. I also show the same SGM plots using a logarithmic scale (with no restrictions on $\vert g(i) - g_0\vert/g_0$). The paths imaged by SGM clearly match with those of the local current density maps. This confirms that the directed paths imaged through the SGM approach used in the paper correspond to the paths taken by the electron flow.

\begin{acknowledgments}
Figure \ref{fig:1DWLlead} and the two upper panels of Fig.~\ref{fig:currvsSGM} have been obtained through the Kwant library \cite{groth2014kwant}. I thank F. Evers for interesting discussions and in particular for suggesting that I compare the local current density with the local response of the conductance considered in the paper.
\end{acknowledgments}


\begin{thebibliography}{86}%
\makeatletter
\providecommand \@ifxundefined [1]{%
 \@ifx{#1\undefined}
}%
\providecommand \@ifnum [1]{%
 \ifnum #1\expandafter \@firstoftwo
 \else \expandafter \@secondoftwo
 \fi
}%
\providecommand \@ifx [1]{%
 \ifx #1\expandafter \@firstoftwo
 \else \expandafter \@secondoftwo
 \fi
}%
\providecommand \natexlab [1]{#1}%
\providecommand \enquote  [1]{``#1''}%
\providecommand \bibnamefont  [1]{#1}%
\providecommand \bibfnamefont [1]{#1}%
\providecommand \citenamefont [1]{#1}%
\providecommand \href@noop [0]{\@secondoftwo}%
\providecommand \href [0]{\begingroup \@sanitize@url \@href}%
\providecommand \@href[1]{\@@startlink{#1}\@@href}%
\providecommand \@@href[1]{\endgroup#1\@@endlink}%
\providecommand \@sanitize@url [0]{\catcode `\\12\catcode `\$12\catcode
  `\&12\catcode `\#12\catcode `\^12\catcode `\_12\catcode `\%12\relax}%
\providecommand \@@startlink[1]{}%
\providecommand \@@endlink[0]{}%
\providecommand \url  [0]{\begingroup\@sanitize@url \@url }%
\providecommand \@url [1]{\endgroup\@href {#1}{\urlprefix }}%
\providecommand \urlprefix  [0]{URL }%
\providecommand \Eprint [0]{\href }%
\providecommand \doibase [0]{http://dx.doi.org/}%
\providecommand \selectlanguage [0]{\@gobble}%
\providecommand \bibinfo  [0]{\@secondoftwo}%
\providecommand \bibfield  [0]{\@secondoftwo}%
\providecommand \translation [1]{[#1]}%
\providecommand \BibitemOpen [0]{}%
\providecommand \bibitemStop [0]{}%
\providecommand \bibitemNoStop [0]{.\EOS\space}%
\providecommand \EOS [0]{\spacefactor3000\relax}%
\providecommand \BibitemShut  [1]{\csname bibitem#1\endcsname}%
\let\auto@bib@innerbib\@empty
%</preamble>
\bibitem [{\citenamefont {Anderson}(1958)}]{Anderson:PR58}%
  \BibitemOpen
  \bibfield  {author} {\bibinfo {author} {\bibfnamefont {P.~W.}\ \bibnamefont
  {Anderson}},\ } {\bibfield
  {journal} {\bibinfo  {journal} {Phys. Rev.}\ }\textbf {\bibinfo {volume}
  {109}},\ \bibinfo {pages} {1492} (\bibinfo {year} {1958})}\BibitemShut
  {NoStop}%
\bibitem [{\citenamefont {Abrahams}(2010)}]{abrahams201050}%
  \BibitemOpen
  \bibfield  {author} {\bibinfo {author} {\bibfnamefont {E.}~\bibnamefont
  {Abrahams}},\ }\href@noop {} {\emph {\bibinfo {title} {50 years of Anderson
  Localization}}}\ (\bibinfo  {publisher} {World Scientific},\ \bibinfo {year}
  {2010})\BibitemShut {NoStop}%
\bibitem [{\citenamefont {Dobrosavljevic}\ \emph {et~al.}(2012)\citenamefont
  {Dobrosavljevic}, \citenamefont {Trivedi},\ and\ \citenamefont
  {Valles~Jr}}]{dobrosavljevic2012conductor}%
  \BibitemOpen
  \bibfield  {author} {\bibinfo {author} {\bibfnamefont {V.}~\bibnamefont
  {Dobrosavljevic}}, \bibinfo {author} {\bibfnamefont {N.}~\bibnamefont
  {Trivedi}}, \ and\ \bibinfo {author} {\bibfnamefont {J.~M.}\ \bibnamefont
  {Valles~Jr}},\ }\href@noop {} {\emph {\bibinfo {title} {Conductor insulator
  quantum phase transitions}}}\ (\bibinfo  {publisher} {Oxford University
  Press},\ \bibinfo {year} {2012})\BibitemShut {NoStop}%
\bibitem [{\citenamefont {Castellani}\ and\ \citenamefont
  {Peliti}(1986)}]{castellani1986multifractal}%
  \BibitemOpen
  \bibfield  {author} {\bibinfo {author} {\bibfnamefont {C.}~\bibnamefont
  {Castellani}}\ and\ \bibinfo {author} {\bibfnamefont {L.}~\bibnamefont
  {Peliti}},\ }\href@noop {} {\bibfield  {journal} {\bibinfo  {journal}
  {J. Phys. A}\ }\textbf {\bibinfo {volume}
  {19}},\ \bibinfo {pages} {L429} (\bibinfo {year} {1986})}\BibitemShut
  {NoStop}%
\bibitem [{\citenamefont {Wegner}(1980)}]{wegner1980inverse}%
  \BibitemOpen
  \bibfield  {author} {\bibinfo {author} {\bibfnamefont {F.}~\bibnamefont
  {Wegner}},\ }\href@noop {} {\bibfield  {journal} {\bibinfo  {journal}
  {Z. Phys. B}\ }\textbf {\bibinfo {volume}
  {36}},\ \bibinfo {pages} {209} (\bibinfo {year} {1980})}\BibitemShut
  {NoStop}%
\bibitem [{\citenamefont {Evers}\ and\ \citenamefont
  {Mirlin}(2008)}]{evers2008anderson}%
  \BibitemOpen
  \bibfield  {author} {\bibinfo {author} {\bibfnamefont {F.}~\bibnamefont
  {Evers}}\ and\ \bibinfo {author} {\bibfnamefont {A.~D.}\ \bibnamefont
  {Mirlin}},\ }\href@noop {} {\bibfield  {journal} {\bibinfo  {journal}
  {Rev. Mod. Phys.}\ }\textbf {\bibinfo {volume} {80}},\ \bibinfo
  {pages} {1355} (\bibinfo {year} {2008})}\BibitemShut {NoStop}%
\bibitem [{\citenamefont {Feigel'Man}\ \emph {et~al.}(2010)\citenamefont
  {Feigel'Man}, \citenamefont {Ioffe}, \citenamefont {Kravtsov},\ and\
  \citenamefont {Cuevas}}]{feigel2010fractal}%
  \BibitemOpen
  \bibfield  {author} {\bibinfo {author} {\bibfnamefont {M.}~\bibnamefont
  {Feigel'Man}}, \bibinfo {author} {\bibfnamefont {L.}~\bibnamefont {Ioffe}},
  \bibinfo {author} {\bibfnamefont {V.}~\bibnamefont {Kravtsov}}, \ and\
  \bibinfo {author} {\bibfnamefont {E.}~\bibnamefont {Cuevas}},\ }\href@noop {}
  {\bibfield  {journal} {\bibinfo  {journal} {Ann. Phys.}\ }\textbf
  {\bibinfo {volume} {325}},\ \bibinfo {pages} {1390} (\bibinfo {year}
  {2010})}\BibitemShut {NoStop}%
\bibitem [{\citenamefont {Altshuler}\ \emph {et~al.}(1997)\citenamefont
  {Altshuler}, \citenamefont {Gefen}, \citenamefont {Kamenev},\ and\
  \citenamefont {Levitov}}]{PhysRevLett.78.2803}%
  \BibitemOpen
  \bibfield  {author} {\bibinfo {author} {\bibfnamefont {B.~L.}\ \bibnamefont
  {Altshuler}}, \bibinfo {author} {\bibfnamefont {Y.}~\bibnamefont {Gefen}},
  \bibinfo {author} {\bibfnamefont {A.}~\bibnamefont {Kamenev}}, \ and\
  \bibinfo {author} {\bibfnamefont {L.~S.}\ \bibnamefont {Levitov}},\ }
  {\bibfield  {journal} {\bibinfo
  {journal} {Phys. Rev. Lett.}\ }\textbf {\bibinfo {volume} {78}},\ \bibinfo
  {pages} {2803} (\bibinfo {year} {1997})}\BibitemShut {NoStop}%
\bibitem [{\citenamefont {Basko}\ \emph {et~al.}(2006)\citenamefont {Basko},
  \citenamefont {Aleiner},\ and\ \citenamefont {Altshuler}}]{basko2006metal}%
  \BibitemOpen
  \bibfield  {author} {\bibinfo {author} {\bibfnamefont {D.}~\bibnamefont
  {Basko}}, \bibinfo {author} {\bibfnamefont {I.}~\bibnamefont {Aleiner}}, \
  and\ \bibinfo {author} {\bibfnamefont {B.}~\bibnamefont {Altshuler}},\
  }\href@noop {} {\bibfield  {journal} {\bibinfo  {journal} {Ann. Phys.}\ }\textbf {\bibinfo {volume} {321}},\ \bibinfo {pages} {1126}
  (\bibinfo {year} {2006})}\BibitemShut {NoStop}%
\bibitem [{\citenamefont {Gornyi}\ \emph {et~al.}(2005)\citenamefont {Gornyi},
  \citenamefont {Mirlin},\ and\ \citenamefont
  {Polyakov}}]{PhysRevLett.95.206603}%
  \BibitemOpen
  \bibfield  {author} {\bibinfo {author} {\bibfnamefont {I.~V.}\ \bibnamefont
  {Gornyi}}, \bibinfo {author} {\bibfnamefont {A.~D.}\ \bibnamefont {Mirlin}},
  \ and\ \bibinfo {author} {\bibfnamefont {D.~G.}\ \bibnamefont {Polyakov}},\
  } {\bibfield  {journal}
  {\bibinfo  {journal} {Phys. Rev. Lett.}\ }\textbf {\bibinfo {volume} {95}},\
  \bibinfo {pages} {206603} (\bibinfo {year} {2005})}\BibitemShut {NoStop}%
\bibitem [{\citenamefont {Nandkishore}\ and\ \citenamefont
  {Huse}(2015)}]{nandkishore2015many}%
  \BibitemOpen
  \bibfield  {author} {\bibinfo {author} {\bibfnamefont {R.}~\bibnamefont
  {Nandkishore}}\ and\ \bibinfo {author} {\bibfnamefont {D.~A.}\ \bibnamefont
  {Huse}},\ }\href@noop {} {\bibfield  {journal} {\bibinfo  {journal} {Annu.
  Rev. Condens. Matter Phys.}\ }\textbf {\bibinfo {volume} {6}},\ \bibinfo
  {pages} {15} (\bibinfo {year} {2015})}\BibitemShut {NoStop}%
\bibitem [{\citenamefont {Abanin}\ \emph {et~al.}(2018)\citenamefont {Abanin},
  \citenamefont {Altman}, \citenamefont {Bloch},\ and\ \citenamefont
  {Serbyn}}]{abanin2018ergodicity}%
  \BibitemOpen
  \bibfield  {author} {\bibinfo {author} {\bibfnamefont {D.~A.}\ \bibnamefont
  {Abanin}}, \bibinfo {author} {\bibfnamefont {E.}~\bibnamefont {Altman}},
  \bibinfo {author} {\bibfnamefont {I.}~\bibnamefont {Bloch}}, \ and\ \bibinfo
  {author} {\bibfnamefont {M.}~\bibnamefont {Serbyn}},\ }\href@noop {}
  {\bibfield  {journal} {\bibinfo  {journal} {arXiv preprint arXiv:1804.11065}\
  } (\bibinfo {year} {2018})}\BibitemShut {NoStop}%
\bibitem [{\citenamefont {M{\'e}zard}\ \emph {et~al.}(1987)\citenamefont
  {M{\'e}zard}, \citenamefont {Parisi},\ and\ \citenamefont
  {Virasoro}}]{mezard1987spin}%
  \BibitemOpen
  \bibfield  {author} {\bibinfo {author} {\bibfnamefont {M.}~\bibnamefont
  {M{\'e}zard}}, \bibinfo {author} {\bibfnamefont {G.}~\bibnamefont {Parisi}},
  \ and\ \bibinfo {author} {\bibfnamefont {M.}~\bibnamefont {Virasoro}},\
  }\href@noop {} {\emph {\bibinfo {title} {Spin glass theory and beyond: An
  Introduction to the Replica Method and Its Applications}}},\ Vol.~\bibinfo
  {volume} {9}\ (\bibinfo  {publisher} {World Scientific Publishing Co Inc},\
  \bibinfo {year} {1987})\BibitemShut {NoStop}%
\bibitem [{\citenamefont {Young}(1998)}]{young1998spin}%
  \BibitemOpen
  \bibfield  {author} {\bibinfo {author} {\bibfnamefont {A.~P.}\ \bibnamefont
  {Young}},\ }\href@noop {} {\emph {\bibinfo {title} {Spin glasses and random
  fields}}},\ Vol.~\bibinfo {volume} {12}\ (\bibinfo  {publisher} {World
  Scientific},\ \bibinfo {year} {1998})\BibitemShut {NoStop}%
\bibitem [{\citenamefont {Stein}\ and\ \citenamefont
  {Newman}(2013)}]{stein2013spin}%
  \BibitemOpen
  \bibfield  {author} {\bibinfo {author} {\bibfnamefont {D.~L.}\ \bibnamefont
  {Stein}}\ and\ \bibinfo {author} {\bibfnamefont {C.~M.}\ \bibnamefont
  {Newman}},\ }\href@noop {} {\emph {\bibinfo {title} {Spin glasses and
  complexity}}}\ (\bibinfo  {publisher} {Princeton University Press},\ \bibinfo
  {year} {2013})\BibitemShut {NoStop}%
\bibitem [{\citenamefont {Davies}\ \emph {et~al.}(1982)\citenamefont {Davies},
  \citenamefont {Lee},\ and\ \citenamefont {Rice}}]{davies1982electron}%
  \BibitemOpen
  \bibfield  {author} {\bibinfo {author} {\bibfnamefont {J.H.}~\bibnamefont
  {Davies}}, \bibinfo {author} {\bibfnamefont {P.A.}~\bibnamefont {Lee}}, \ and\
  \bibinfo {author} {\bibfnamefont {T.M.}~\bibnamefont {Rice}},\ }\href@noop {}
  {\bibfield  {journal} {\bibinfo  {journal} {Phys. Rev. Lett.}\
  }\textbf {\bibinfo {volume} {49}},\ \bibinfo {pages} {758} (\bibinfo {year}
  {1982})}\BibitemShut {NoStop}%
\bibitem [{\citenamefont {M\"uller}\ and\ \citenamefont
  {Ioffe}(2004)}]{PhysRevLett.93.256403}%
  \BibitemOpen
  \bibfield  {author} {\bibinfo {author} {\bibfnamefont {M.}~\bibnamefont
  {M\"uller}}\ and\ \bibinfo {author} {\bibfnamefont {L.~B.}\ \bibnamefont
  {Ioffe}},\ } {\bibfield
  {journal} {\bibinfo  {journal} {Phys. Rev. Lett.}\ }\textbf {\bibinfo
  {volume} {93}},\ \bibinfo {pages} {256403} (\bibinfo {year}
  {2004})}\BibitemShut {NoStop}%
\bibitem [{\citenamefont {Somoza}\ \emph {et~al.}(2008)\citenamefont {Somoza},
  \citenamefont {Ortu\~no}, \citenamefont {Caravaca},\ and\ \citenamefont
  {Pollak}}]{PhysRevLett.101.056601}%
  \BibitemOpen
  \bibfield  {author} {\bibinfo {author} {\bibfnamefont {A.~M.}\ \bibnamefont
  {Somoza}}, \bibinfo {author} {\bibfnamefont {M.}~\bibnamefont {Ortu\~no}},
  \bibinfo {author} {\bibfnamefont {M.}~\bibnamefont {Caravaca}}, \ and\
  \bibinfo {author} {\bibfnamefont {M.}~\bibnamefont {Pollak}},\ }
  {\bibfield  {journal} {\bibinfo
  {journal} {Phys. Rev. Lett.}\ }\textbf {\bibinfo {volume} {101}},\ \bibinfo
  {pages} {056601} (\bibinfo {year} {2008})}\BibitemShut {NoStop}%
\bibitem [{\citenamefont {Vaknin}\ \emph {et~al.}(2000)\citenamefont {Vaknin},
  \citenamefont {Ovadyahu},\ and\ \citenamefont
  {Pollak}}]{PhysRevLett.84.3402}%
  \BibitemOpen
  \bibfield  {author} {\bibinfo {author} {\bibfnamefont {A.}~\bibnamefont
  {Vaknin}}, \bibinfo {author} {\bibfnamefont {Z.}~\bibnamefont {Ovadyahu}}, \
  and\ \bibinfo {author} {\bibfnamefont {M.}~\bibnamefont {Pollak}},\ }
  {\bibfield  {journal} {\bibinfo
  {journal} {Phys. Rev. Lett.}\ }\textbf {\bibinfo {volume} {84}},\ \bibinfo
  {pages} {3402} (\bibinfo {year} {2000})}\BibitemShut {NoStop}%
\bibitem [{\citenamefont {Amir}\ \emph {et~al.}(2011)\citenamefont {Amir},
  \citenamefont {Oreg},\ and\ \citenamefont {Imry}}]{amir2011electron}%
  \BibitemOpen
  \bibfield  {author} {\bibinfo {author} {\bibfnamefont {A.}~\bibnamefont
  {Amir}}, \bibinfo {author} {\bibfnamefont {Y.}~\bibnamefont {Oreg}}, \ and\
  \bibinfo {author} {\bibfnamefont {Y.}~\bibnamefont {Imry}},\ }\href@noop {}
  {\bibfield  {journal} {\bibinfo  {journal} {Annu. Rev. Condens. Matter
  Phys.}\ }\textbf {\bibinfo {volume} {2}},\ \bibinfo {pages} {235} (\bibinfo
  {year} {2011})}\BibitemShut {NoStop}%
\bibitem [{\citenamefont {Pollak}\ and\ \citenamefont
  {Shklovskii}(1991)}]{pollak1991hopping}%
  \BibitemOpen
  \bibfield  {author} {\bibinfo {author} {\bibfnamefont {M.}~\bibnamefont
  {Pollak}}\ and\ \bibinfo {author} {\bibfnamefont {B.}~\bibnamefont
  {Shklovskii}},\ }\href@noop {} {\emph {\bibinfo {title} {Hopping transport in
  solids}}},\ Vol.~\bibinfo {volume} {28}\ (\bibinfo  {publisher} {Elsevier},\
  \bibinfo {year} {1991})\BibitemShut {NoStop}%
\bibitem [{\citenamefont {Monthus}\ and\ \citenamefont
  {Garel}(2008)}]{monthus2008anderson}%
  \BibitemOpen
  \bibfield  {author} {\bibinfo {author} {\bibfnamefont {C.}~\bibnamefont
  {Monthus}}\ and\ \bibinfo {author} {\bibfnamefont {T.}~\bibnamefont
  {Garel}},\ }\href@noop {} {\bibfield  {journal} {\bibinfo  {journal} {J.
  Phys. A}\ }\textbf {\bibinfo {volume}
  {42}},\ \bibinfo {pages} {075002} (\bibinfo {year} {2008})}\BibitemShut
  {NoStop}%
\bibitem [{\citenamefont {Biroli}\ \emph {et~al.}(2012)\citenamefont {Biroli},
  \citenamefont {Ribeiro-Teixeira},\ and\ \citenamefont
  {Tarzia}}]{biroli2012difference}%
  \BibitemOpen
  \bibfield  {author} {\bibinfo {author} {\bibfnamefont {G.}~\bibnamefont
  {Biroli}}, \bibinfo {author} {\bibfnamefont {A.}~\bibnamefont
  {Ribeiro-Teixeira}}, \ and\ \bibinfo {author} {\bibfnamefont
  {M.}~\bibnamefont {Tarzia}},\ }\href@noop {} {\bibfield  {journal} {\bibinfo
  {journal} {arXiv preprint arXiv:1211.7334}\ } (\bibinfo {year}
  {2012})}\BibitemShut {NoStop}%
\bibitem [{\citenamefont {De~Luca}\ \emph {et~al.}(2014)\citenamefont
  {De~Luca}, \citenamefont {Altshuler}, \citenamefont {Kravtsov},\ and\
  \citenamefont {Scardicchio}}]{deluca2014anderson}%
  \BibitemOpen
  \bibfield  {author} {\bibinfo {author} {\bibfnamefont {A.}~\bibnamefont
  {De~Luca}}, \bibinfo {author} {\bibfnamefont {B.L.}~\bibnamefont {Altshuler}}, \bibinfo {author} {\bibfnamefont {V.E.}~\bibnamefont
  {Kravtsov}}, \ and\ \bibinfo {author} {\bibfnamefont
  {A.}~\bibnamefont {Scardicchio}},\ }\href@noop {} {\bibfield  {journal}
  {\bibinfo  {journal} {Phys. Rev. Lett.}\ }\textbf {\bibinfo {volume} {113}},\
  \bibinfo {pages} {046806} (\bibinfo {year} {2014})}\BibitemShut {NoStop}%
\bibitem [{\citenamefont {Kravtsov}\ \emph {et~al.}(2015)\citenamefont
  {Kravtsov}, \citenamefont {Khaymovich}, \citenamefont {Cuevas},\ and\
  \citenamefont {Amini}}]{kravtsov2015random}%
  \BibitemOpen
  \bibfield  {author} {\bibinfo {author} {\bibfnamefont {V.E.}~\bibnamefont
  {Kravtsov}}, \bibinfo {author} {\bibfnamefont {I.}~\bibnamefont
  {Khaymovich}}, \bibinfo {author} {\bibfnamefont {E.}~\bibnamefont {Cuevas}},
  \ and\ \bibinfo {author} {\bibfnamefont {M.}~\bibnamefont {Amini}},\
  }\href@noop {} {\bibfield  {journal} {\bibinfo  {journal} {New J.
  Phys.}\ }\textbf {\bibinfo {volume} {17}},\ \bibinfo {pages} {122002}
  (\bibinfo {year} {2015})}\BibitemShut {NoStop}%
\bibitem [{\citenamefont {Tikhonov}\ \emph {et~al.}(2016)\citenamefont
  {Tikhonov}, \citenamefont {Mirlin},\ and\ \citenamefont
  {Skvortsov}}]{tikhonov2016anderson}%
  \BibitemOpen
  \bibfield  {author} {\bibinfo {author} {\bibfnamefont {K.S.}~\bibnamefont
  {Tikhonov}}, \bibinfo {author} {\bibfnamefont {A.D.}~\bibnamefont {Mirlin}}, \
  and\ \bibinfo {author} {\bibfnamefont {M.A.}~\bibnamefont {Skvortsov}},\
  }\href@noop {} {\bibfield  {journal} {\bibinfo  {journal} {Phys. Rev.
  B}\ }\textbf {\bibinfo {volume} {94}},\ \bibinfo {pages} {220203} (\bibinfo
  {year} {2016})}\BibitemShut {NoStop}%
\bibitem [{\citenamefont {Garc\'{\i}a-Mata}\ \emph {et~al.}(2017)\citenamefont
  {Garc\'{\i}a-Mata}, \citenamefont {Giraud}, \citenamefont {Georgeot},
  \citenamefont {Martin}, \citenamefont {Dubertrand},\ and\ \citenamefont
  {Lemari\'e}}]{PhysRevLett.118.166801}%
  \BibitemOpen
  \bibfield  {author} {\bibinfo {author} {\bibfnamefont {I.}~\bibnamefont
  {Garc\'{\i}a-Mata}}, \bibinfo {author} {\bibfnamefont {O.}~\bibnamefont
  {Giraud}}, \bibinfo {author} {\bibfnamefont {B.}~\bibnamefont {Georgeot}},
  \bibinfo {author} {\bibfnamefont {J.}~\bibnamefont {Martin}}, \bibinfo
  {author} {\bibfnamefont {R.}~\bibnamefont {Dubertrand}}, \ and\ \bibinfo
  {author} {\bibfnamefont {G.}~\bibnamefont {Lemari\'e}},\ }\href@noop {}
  {\bibfield  {journal} {\bibinfo  {journal} {Phys. Rev. Lett.}\ }\textbf
  {\bibinfo {volume} {118}},\ \bibinfo {pages} {166801} (\bibinfo {year}
  {2017})}\BibitemShut {NoStop}%
\bibitem [{\citenamefont {Biroli}\ and\ \citenamefont
  {Tarzia}(2017)}]{biroli2017delocalized}%
  \BibitemOpen
  \bibfield  {author} {\bibinfo {author} {\bibfnamefont {G.}~\bibnamefont
  {Biroli}}\ and\ \bibinfo {author} {\bibfnamefont {M.}~\bibnamefont
  {Tarzia}},\ }\href@noop {} {\bibfield  {journal} {\bibinfo  {journal} {arXiv
  preprint arXiv:1706.02655}\ } (\bibinfo {year} {2017})}\BibitemShut {NoStop}%
\bibitem [{\citenamefont {Tikhonov}\ and\ \citenamefont
  {Mirlin}(2016)}]{tikhonov2016fractality}%
  \BibitemOpen
  \bibfield  {author} {\bibinfo {author} {\bibfnamefont {K.S.}~\bibnamefont
  {Tikhonov}}\ and\ \bibinfo {author} {\bibfnamefont {A.D.}~\bibnamefont
  {Mirlin}},\ }\href@noop {} {\bibfield  {journal} {\bibinfo  {journal}
  {Phys. Rev. B}\ }\textbf {\bibinfo {volume} {94}},\ \bibinfo {pages}
  {184203} (\bibinfo {year} {2016})}\BibitemShut {NoStop}%
\bibitem [{\citenamefont {Facoetti}\ \emph {et~al.}(2016)\citenamefont
  {Facoetti}, \citenamefont {Vivo},\ and\ \citenamefont
  {Biroli}}]{facoetti2016non}%
  \BibitemOpen
  \bibfield  {author} {\bibinfo {author} {\bibfnamefont {D.}~\bibnamefont
  {Facoetti}}, \bibinfo {author} {\bibfnamefont {P.}~\bibnamefont {Vivo}}, \
  and\ \bibinfo {author} {\bibfnamefont {G.}~\bibnamefont {Biroli}},\
  }\href@noop {} {\bibfield  {journal} {\bibinfo  {journal} {Europhys.
  Lett.}\ }\textbf {\bibinfo {volume} {115}},\ \bibinfo {pages} {47003}
  (\bibinfo {year} {2016})}\BibitemShut {NoStop}%
\bibitem [{\citenamefont {Kravtsov}\ \emph {et~al.}(2017)\citenamefont
  {Kravtsov}, \citenamefont {Altshuler},\ and\ \citenamefont
  {Ioffe}}]{kravtsov2017non}%
  \BibitemOpen
  \bibfield  {author} {\bibinfo {author} {\bibfnamefont {V.}~\bibnamefont
  {Kravtsov}}, \bibinfo {author} {\bibfnamefont {B.}~\bibnamefont {Altshuler}},
  \ and\ \bibinfo {author} {\bibfnamefont {L.}~\bibnamefont {Ioffe}},\
  }\href@noop {} {\bibfield  {journal} {\bibinfo  {journal} {Ann.
  Phys.}\ } (\bibinfo {year} {2017})}\BibitemShut {NoStop}%
\bibitem [{\citenamefont {Sonner}\ \emph {et~al.}(2017)\citenamefont {Sonner},
  \citenamefont {Tikhonov},\ and\ \citenamefont
  {Mirlin}}]{sonner2017multifractality}%
  \BibitemOpen
  \bibfield  {author} {\bibinfo {author} {\bibfnamefont {M.}~\bibnamefont
  {Sonner}}, \bibinfo {author} {\bibfnamefont {K.S.}~\bibnamefont {Tikhonov}}, \
  and\ \bibinfo {author} {\bibfnamefont {A.D.}~\bibnamefont {Mirlin}},\
  }\href@noop {} {\bibfield  {journal} {\bibinfo  {journal} {Phys. Rev.
  B}\ }\textbf {\bibinfo {volume} {96}},\ \bibinfo {pages} {214204} (\bibinfo
  {year} {2017})}\BibitemShut {NoStop}%
\bibitem [{\citenamefont {Nguyen}\ \emph {et~al.}(1985)\citenamefont {Nguyen},
  \citenamefont {Spivak},\ and\ \citenamefont {Shklovskii}}]{nguyen1985tunnel}%
  \BibitemOpen
  \bibfield  {author} {\bibinfo {author} {\bibfnamefont {V.}~\bibnamefont
  {Nguyen}}, \bibinfo {author} {\bibfnamefont {B.}~\bibnamefont {Spivak}}, \
  and\ \bibinfo {author} {\bibfnamefont {B.}~\bibnamefont {Shklovskii}},\
  }\href@noop {} {\bibfield  {journal} {\bibinfo  {journal} {Sov. Phys. JETP}\
  }\textbf {\bibinfo {volume} {62}},\ \bibinfo {pages} {1021} (\bibinfo {year}
  {1985})}\BibitemShut {NoStop}%
\bibitem [{\citenamefont {Fritzsche}\ and\ \citenamefont
  {Pollak}(1990)}]{fritzsche1990hopping}%
  \BibitemOpen
  \bibfield  {author} {\bibinfo {author} {\bibfnamefont {H.}~\bibnamefont
  {Fritzsche}}\ and\ \bibinfo {author} {\bibfnamefont {M.}~\bibnamefont
  {Pollak}},\ }\href@noop {} {\emph {\bibinfo {title} {Hopping and related
  phenomena}}},\ Vol.~\bibinfo {volume} {2}\ (\bibinfo  {publisher} {World
  Scientific},\ \bibinfo {year} {1990})\BibitemShut {NoStop}%
\bibitem [{\citenamefont {Medina}\ \emph {et~al.}(1989)\citenamefont {Medina},
  \citenamefont {Kardar}, \citenamefont {Shapir},\ and\ \citenamefont
  {Wang}}]{medina1989interference}%
  \BibitemOpen
  \bibfield  {author} {\bibinfo {author} {\bibfnamefont {E.}~\bibnamefont
  {Medina}}, \bibinfo {author} {\bibfnamefont {M.}~\bibnamefont {Kardar}},
  \bibinfo {author} {\bibfnamefont {Y.}~\bibnamefont {Shapir}}, \ and\ \bibinfo
  {author} {\bibfnamefont {X.~R.}\ \bibnamefont {Wang}},\ }\href@noop {}
  {\bibfield  {journal} {\bibinfo  {journal} {Phys. Rev. Lett.}\
  }\textbf {\bibinfo {volume} {62}},\ \bibinfo {pages} {941} (\bibinfo {year}
  {1989})}\BibitemShut {NoStop}%
\bibitem [{\citenamefont {Pietracaprina}\ \emph {et~al.}(2016)\citenamefont
  {Pietracaprina}, \citenamefont {Ros},\ and\ \citenamefont
  {Scardicchio}}]{PhysRevB.93.054201}%
  \BibitemOpen
  \bibfield  {author} {\bibinfo {author} {\bibfnamefont {F.}~\bibnamefont
  {Pietracaprina}}, \bibinfo {author} {\bibfnamefont {V.}~\bibnamefont {Ros}},
  \ and\ \bibinfo {author} {\bibfnamefont {A.}~\bibnamefont {Scardicchio}},\
  } {\bibfield  {journal} {\bibinfo
  {journal} {Phys. Rev. B}\ }\textbf {\bibinfo {volume} {93}},\ \bibinfo
  {pages} {054201} (\bibinfo {year} {2016})}\BibitemShut {NoStop}%
\bibitem [{\citenamefont {Medina}\ and\ \citenamefont
  {Kardar}(1992)}]{PhysRevB.46.9984}%
  \BibitemOpen
  \bibfield  {author} {\bibinfo {author} {\bibfnamefont {E.}~\bibnamefont
  {Medina}}\ and\ \bibinfo {author} {\bibfnamefont {M.}~\bibnamefont
  {Kardar}},\ } {\bibfield  {journal}
  {\bibinfo  {journal} {Phys. Rev. B}\ }\textbf {\bibinfo {volume} {46}},\
  \bibinfo {pages} {9984} (\bibinfo {year} {1992})}\BibitemShut {NoStop}%
  \bibitem [{\citenamefont {Halpin-Healy}\ and\ \citenamefont
  {Zhang}(1995)}]{halpin1995kinetic}%
  \BibitemOpen
  \bibfield  {author} {\bibinfo {author} {\bibfnamefont {T.}~\bibnamefont
  {Halpin-Healy}}\ and\ \bibinfo {author} {\bibfnamefont {Y.-C.}\ \bibnamefont
  {Zhang}},\ }\href@noop {} {\bibfield  {journal} {\bibinfo  {journal} {Phys.
  Rep.}\ }\textbf {\bibinfo {volume} {254}},\ \bibinfo {pages} {215}
  (\bibinfo {year} {1995})}\BibitemShut {NoStop}%
\bibitem [{\citenamefont {Prior}\ \emph {et~al.}(2005)\citenamefont {Prior},
  \citenamefont {Somoza},\ and\ \citenamefont {Ortu\~no}}]{prior2005conductance}%
  \BibitemOpen
  \bibfield  {author} {\bibinfo {author} {\bibfnamefont {J.}~\bibnamefont
  {Prior}}, \bibinfo {author} {\bibfnamefont {A.M.}~\bibnamefont {Somoza}}, \
  and\ \bibinfo {author} {\bibfnamefont {M.}~\bibnamefont {Ortu\~no}},\
  }\href@noop {} {\bibfield  {journal} {\bibinfo  {journal} {Phys. Rev.
  B}\ }\textbf {\bibinfo {volume} {72}},\ \bibinfo {pages} {024206} (\bibinfo
  {year} {2005})}\BibitemShut {NoStop}%
\bibitem [{\citenamefont {Somoza}\ \emph {et~al.}(2007)\citenamefont {Somoza},
  \citenamefont {Ortu\~no},\ and\ \citenamefont
  {Prior}}]{PhysRevLett.99.116602}%
  \BibitemOpen
  \bibfield  {author} {\bibinfo {author} {\bibfnamefont {A.~M.}\ \bibnamefont
  {Somoza}}, \bibinfo {author} {\bibfnamefont {M.}~\bibnamefont {Ortu\~no}}, \
  and\ \bibinfo {author} {\bibfnamefont {J.}~\bibnamefont {Prior}},\ }
  {\bibfield  {journal} {\bibinfo
  {journal} {Phys. Rev. Lett.}\ }\textbf {\bibinfo {volume} {99}},\ \bibinfo
  {pages} {116602} (\bibinfo {year} {2007})}\BibitemShut {NoStop}%
\bibitem [{\citenamefont {Prior}\ \emph {et~al.}(2009)\citenamefont {Prior},
  \citenamefont {Somoza},\ and\ \citenamefont {Ortu\~no}}]{prior2009conductance}%
  \BibitemOpen
  \bibfield  {author} {\bibinfo {author} {\bibfnamefont {J.}~\bibnamefont
  {Prior}}, \bibinfo {author} {\bibfnamefont {A.}~\bibnamefont {Somoza}}, \
  and\ \bibinfo {author} {\bibfnamefont {M.}~\bibnamefont {Ortu\~no}},\
  }\href@noop {} {\bibfield  {journal} {\bibinfo  {journal} {Eur.
  Phys. J. B}\ }\textbf {\bibinfo {volume} {70}},\ \bibinfo {pages}
  {513} (\bibinfo {year} {2009})}\BibitemShut {NoStop}%
  \bibitem [{\citenamefont {Somoza}\ \emph {et~al.}(2015)\citenamefont {Somoza},
  \citenamefont {Le~Doussal},\ and\ \citenamefont
  {Ortu\~no}}]{PhysRevB.91.155413}%
  \BibitemOpen
  \bibfield  {author} {\bibinfo {author} {\bibfnamefont {A.~M.}\ \bibnamefont
  {Somoza}}, \bibinfo {author} {\bibfnamefont {P.}~\bibnamefont {Le~Doussal}},
  \ and\ \bibinfo {author} {\bibfnamefont {M.}~\bibnamefont {Ortu\~no}},\
  } {\bibfield  {journal} {\bibinfo
  {journal} {Phys. Rev. B}\ }\textbf {\bibinfo {volume} {91}},\ \bibinfo
  {pages} {155413} (\bibinfo {year} {2015})}\BibitemShut {NoStop}%
\bibitem [{\citenamefont {Derrida}\ and\ \citenamefont
  {Spohn}(1988)}]{derrida1988polymers}%
  \BibitemOpen
  \bibfield  {author} {\bibinfo {author} {\bibfnamefont {B.}~\bibnamefont
  {Derrida}}\ and\ \bibinfo {author} {\bibfnamefont {H.}~\bibnamefont
  {Spohn}},\ }\href@noop {} {\bibfield  {journal} {\bibinfo  {journal} {J.
  Stat. Phys.}\ }\textbf {\bibinfo {volume} {51}},\ \bibinfo {pages}
  {817} (\bibinfo {year} {1988})}\BibitemShut {NoStop}%
\bibitem [{\citenamefont {M{\'e}zard}(1990)}]{mezard1990glassy}%
  \BibitemOpen
  \bibfield  {author} {\bibinfo {author} {\bibfnamefont {M.}~\bibnamefont
  {M{\'e}zard}},\ }\href@noop {} {\bibfield  {journal} {\bibinfo  {journal}
  {J. Phys. (Paris)}\ }\textbf {\bibinfo {volume} {51}},\ \bibinfo {pages}
  {1831} (\bibinfo {year} {1990})}\BibitemShut {NoStop}%
\bibitem [{\citenamefont {J\"ogi}\ and\ \citenamefont
  {Sornette}(1998)}]{PhysRevE.57.6936}%
  \BibitemOpen
  \bibfield  {author} {\bibinfo {author} {\bibfnamefont {P.}~\bibnamefont
  {J\"ogi}}\ and\ \bibinfo {author} {\bibfnamefont {D.}~\bibnamefont
  {Sornette}},\ }{\bibfield
  {journal} {\bibinfo  {journal} {Phys. Rev. E}\ }\textbf {\bibinfo {volume}
  {57}},\ \bibinfo {pages} {6936} (\bibinfo {year} {1998})}\BibitemShut
  {NoStop}%
\bibitem [{\citenamefont {Thiery}(2017)}]{thiery2017analytical}%
  \BibitemOpen
  \bibfield  {author} {\bibinfo {author} {\bibfnamefont {T.}~\bibnamefont
  {Thiery}},\ }\href@noop {} {\bibfield  {journal} {\bibinfo  {journal} {arXiv
  preprint arXiv:1705.07457}\ } (\bibinfo {year} {2017})}\BibitemShut {NoStop}%
\bibitem [{\citenamefont {McKay}\ \emph {et~al.}(1982)\citenamefont {McKay},
  \citenamefont {Berker},\ and\ \citenamefont {Kirkpatrick}}]{mckay1982spin}%
  \BibitemOpen
  \bibfield  {author} {\bibinfo {author} {\bibfnamefont {S.~R.}\ \bibnamefont
  {McKay}}, \bibinfo {author} {\bibfnamefont {A.~N.}\ \bibnamefont {Berker}}, \
  and\ \bibinfo {author} {\bibfnamefont {S.}~\bibnamefont {Kirkpatrick}},\
  }\href@noop {} {\bibfield  {journal} {\bibinfo  {journal} {Phys. Rev.
  Lett.}\ }\textbf {\bibinfo {volume} {48}},\ \bibinfo {pages} {767}
  (\bibinfo {year} {1982})}\BibitemShut {NoStop}%
\bibitem [{\citenamefont {Bray}\ and\ \citenamefont
  {Moore}(1987)}]{PhysRevLett.58.57}%
  \BibitemOpen
  \bibfield  {author} {\bibinfo {author} {\bibfnamefont {A.~J.}\ \bibnamefont
  {Bray}}\ and\ \bibinfo {author} {\bibfnamefont {M.~A.}\ \bibnamefont
  {Moore}},\ } {\bibfield  {journal}
  {\bibinfo  {journal} {Phys. Rev. Lett.}\ }\textbf {\bibinfo {volume} {58}},\
  \bibinfo {pages} {57} (\bibinfo {year} {1987})}\BibitemShut {NoStop}%
\bibitem [{\citenamefont {Fisher}\ and\ \citenamefont
  {Huse}(1988)}]{PhysRevB.38.386}%
  \BibitemOpen
  \bibfield  {author} {\bibinfo {author} {\bibfnamefont {D.~S.}\ \bibnamefont
  {Fisher}}\ and\ \bibinfo {author} {\bibfnamefont {D.~A.}\ \bibnamefont
  {Huse}},\ }{\bibfield  {journal}
  {\bibinfo  {journal} {Phys. Rev. B}\ }\textbf {\bibinfo {volume} {38}},\
  \bibinfo {pages} {386} (\bibinfo {year} {1988})}\BibitemShut {NoStop}%
\bibitem [{\citenamefont {Fisher}\ and\ \citenamefont
  {Huse}(1991)}]{PhysRevB.43.10728}%
  \BibitemOpen
  \bibfield  {author} {\bibinfo {author} {\bibfnamefont {D.~S.}\ \bibnamefont
  {Fisher}}\ and\ \bibinfo {author} {\bibfnamefont {D.~A.}\ \bibnamefont
  {Huse}},\ } {\bibfield  {journal}
  {\bibinfo  {journal} {Phys. Rev. B}\ }\textbf {\bibinfo {volume} {43}},\
  \bibinfo {pages} {10728} (\bibinfo {year} {1991})}\BibitemShut {NoStop}%
\bibitem [{\citenamefont {Sales}\ and\ \citenamefont
  {Yoshino}(2002)}]{sales2002fragility}%
  \BibitemOpen
  \bibfield  {author} {\bibinfo {author} {\bibfnamefont {M.}~\bibnamefont
  {Sales}}\ and\ \bibinfo {author} {\bibfnamefont {H.}~\bibnamefont
  {Yoshino}},\ }\href@noop {} {\bibfield  {journal} {\bibinfo  {journal}
  {Phys. Rev. E}\ }\textbf {\bibinfo {volume} {65}},\ \bibinfo {pages}
  {066131} (\bibinfo {year} {2002})}\BibitemShut {NoStop}%
\bibitem [{\citenamefont {da~Silveira}\ and\ \citenamefont
  {Bouchaud}(2004)}]{da2004temperature}%
  \BibitemOpen
  \bibfield  {author} {\bibinfo {author} {\bibfnamefont {R.~A.}\ \bibnamefont
  {da~Silveira}}\ and\ \bibinfo {author} {\bibfnamefont {J.-P.}\ \bibnamefont
  {Bouchaud}},\ }\href@noop {} {\bibfield  {journal} {\bibinfo  {journal}
  {Phys. Rev. Lett.}\ }\textbf {\bibinfo {volume} {93}},\ \bibinfo
  {pages} {015901} (\bibinfo {year} {2004})}\BibitemShut {NoStop}%
\bibitem{PhysRevLett.96.235702}%
  \BibitemOpen
  \bibfield  {author} {\bibinfo {author} {\bibfnamefont {P.}\ \bibnamefont
  {Le~Doussal}},\ }\href@noop {} {\bibfield  {journal} {\bibinfo  {journal}
  {Phys. Rev. Lett.}\ }\textbf {\bibinfo {volume} {96}},\ \bibinfo
  {pages} {235702} (\bibinfo {year} {2006})}\BibitemShut {NoStop}%
\bibitem [{\citenamefont {Balents}\ \emph {et~al.}(1996)\citenamefont
  {Balents}, \citenamefont {Bouchaud},\ and\ \citenamefont
  {M{\'e}zard}}]{balents1996large}%
  \BibitemOpen
  \bibfield  {author} {\bibinfo {author} {\bibfnamefont {L.}~\bibnamefont
  {Balents}}, \bibinfo {author} {\bibfnamefont {J.-P.}\ \bibnamefont
  {Bouchaud}}, \ and\ \bibinfo {author} {\bibfnamefont {M.}~\bibnamefont
  {M{\'e}zard}},\ }\href@noop {} {\bibfield  {journal} {\bibinfo  {journal}
  {J. Phys. I}\ }\textbf {\bibinfo {volume} {6}},\ \bibinfo {pages}
  {1007} (\bibinfo {year} {1996})}\BibitemShut {NoStop}%
\bibitem [{\citenamefont {Datta}(1997)}]{datta1997electronic}%
  \BibitemOpen
  \bibfield  {author} {\bibinfo {author} {\bibfnamefont {S.}~\bibnamefont
  {Datta}},\ }\href@noop {} {\emph {\bibinfo {title} {Electronic transport in
  mesoscopic systems}}}\ (\bibinfo  {publisher} {Cambridge university press},\
  \bibinfo {year} {1997})\BibitemShut {NoStop}%
  \bibitem [{\citenamefont {Abbout}\ \emph {et~al.}(2011)\citenamefont {Abbout},
  \citenamefont {Lemari{\'e}},\ and\ \citenamefont
  {Pichard}}]{abbout2011thermal}%
  \BibitemOpen
  \bibfield  {author} {\bibinfo {author} {\bibfnamefont {A.}~\bibnamefont
  {Abbout}}, \bibinfo {author} {\bibfnamefont {G.}~\bibnamefont {Lemari{\'e}}},
  \ and\ \bibinfo {author} {\bibfnamefont {J.-L.}\ \bibnamefont {Pichard}},\
  }\href@noop {} {\bibfield  {journal} {\bibinfo  {journal} {Phys. Rev.
  Lett.}\ }\textbf {\bibinfo {volume} {106}},\ \bibinfo {pages} {156810}
  (\bibinfo {year} {2011})}\BibitemShut {NoStop}%
  \bibitem{pichard1991quantum}%
J.-L. Pichard, in \textit{Quantum Coherence in Mesoscopic Systems}, edited by B. Kramer, NATO ASI Series B254 (Plenum,
New York) (1991).
    \bibitem [{\citenamefont {Marko{\v{s}}}(2010)}]{markovs2010electron}%
  \BibitemOpen
  \bibfield  {author} {\bibinfo {author} {\bibfnamefont {P.}~\bibnamefont
  {Marko{\v{s}}}},\ }\href@noop {} {\bibfield  {journal} {\bibinfo  {journal}
  {Physica B}\ }\textbf {\bibinfo {volume} {405}},\ \bibinfo
  {pages} {3029} (\bibinfo {year} {2010})}\BibitemShut {NoStop}%
  \bibitem [{\citenamefont {Topinka}\ \emph
  {et~al.}(2001{\natexlab{b}})\citenamefont {Topinka}, \citenamefont {LeRoy},
  \citenamefont {Westervelt}, \citenamefont {Shaw}, \citenamefont
  {Fleischmann}, \citenamefont {Heller}, \citenamefont {Maranowski},\ and\
  \citenamefont {Gossard}}]{topinka2001coherent}%
  \BibitemOpen
  \bibfield  {author} {\bibinfo {author} {\bibfnamefont {M.}~\bibnamefont
  {Topinka}}, \bibinfo {author} {\bibfnamefont {B.}~\bibnamefont {LeRoy}},
  \bibinfo {author} {\bibfnamefont {R.}~\bibnamefont {Westervelt}}, \bibinfo
  {author} {\bibfnamefont {S.}~\bibnamefont {Shaw}}, \bibinfo {author}
  {\bibfnamefont {R.}~\bibnamefont {Fleischmann}}, \bibinfo {author}
  {\bibfnamefont {E.}~\bibnamefont {Heller}}, \bibinfo {author} {\bibfnamefont
  {K.}~\bibnamefont {Maranowski}}, \ and\ \bibinfo {author} {\bibfnamefont
  {A.}~\bibnamefont {Gossard}},\ }\href@noop {} {\bibfield  {journal} {\bibinfo
   {journal} {Nature}\ }\textbf {\bibinfo {volume} {410}},\ \bibinfo {pages}
  {183} (\bibinfo {year} {2001}{\natexlab{b}})}\BibitemShut {NoStop}%
\bibitem [{\citenamefont {Gorini}\ \emph {et~al.}(2013)\citenamefont {Gorini},
  \citenamefont {Jalabert}, \citenamefont {Szewc}, \citenamefont {Tomsovic},\
  and\ \citenamefont {Weinmann}}]{gorini2013theory}%
  \BibitemOpen
  \bibfield  {author} {\bibinfo {author} {\bibfnamefont {C.}~\bibnamefont
  {Gorini}}, \bibinfo {author} {\bibfnamefont {R.~A.}\ \bibnamefont
  {Jalabert}}, \bibinfo {author} {\bibfnamefont {W.}~\bibnamefont {Szewc}},
  \bibinfo {author} {\bibfnamefont {S.}~\bibnamefont {Tomsovic}}, \ and\
  \bibinfo {author} {\bibfnamefont {D.}~\bibnamefont {Weinmann}},\ }\href@noop
  {} {\bibfield  {journal} {\bibinfo  {journal} {Physical Review B}\ }\textbf
  {\bibinfo {volume} {88}},\ \bibinfo {pages} {035406} (\bibinfo {year}
  {2013})}\BibitemShut {NoStop}%
  \bibitem{supmat} See Supplemental Material. \BibitemShut {NoStop}% 
\bibitem [{\citenamefont {Fisher}\ and\ \citenamefont
  {Lee}(1981)}]{fisher1981relation}%
  \BibitemOpen
  \bibfield  {author} {\bibinfo {author} {\bibfnamefont {D.~S.}\ \bibnamefont
  {Fisher}}\ and\ \bibinfo {author} {\bibfnamefont {P.~A.}\ \bibnamefont
  {Lee}},\ }\href@noop {} {\bibfield  {journal} {\bibinfo  {journal} {Phys.
  Rev. B}\ }\textbf {\bibinfo {volume} {23}},\ \bibinfo {pages} {6851}
  (\bibinfo {year} {1981})}\BibitemShut {NoStop}%
\bibitem [{\citenamefont {Gueudre}\ \emph {et~al.}(2015)\citenamefont
  {Gueudre}, \citenamefont {Le~Doussal}, \citenamefont {Bouchaud},\ and\
  \citenamefont {Rosso}}]{gueudre2015ground}%
  \BibitemOpen
  \bibfield  {author} {\bibinfo {author} {\bibfnamefont {T.}~\bibnamefont
  {Gueudre}}, \bibinfo {author} {\bibfnamefont {P.}~\bibnamefont {Le~Doussal}},
  \bibinfo {author} {\bibfnamefont {J.-P.}\ \bibnamefont {Bouchaud}}, \ and\
  \bibinfo {author} {\bibfnamefont {A.}~\bibnamefont {Rosso}},\ }\href@noop {}
  {\bibfield  {journal} {\bibinfo  {journal} {Phys. Rev. E}\ }\textbf
  {\bibinfo {volume} {91}},\ \bibinfo {pages} {062110} (\bibinfo {year}
  {2015})}\BibitemShut {NoStop}%
\bibitem [{\citenamefont {Maes}\ and\ \citenamefont
  {Thiery}(2017)}]{maes2017midpoint}%
  \BibitemOpen
  \bibfield  {author} {\bibinfo {author} {\bibfnamefont {C.}~\bibnamefont
  {Maes}}\ and\ \bibinfo {author} {\bibfnamefont {T.}~\bibnamefont {Thiery}},\
  }\href@noop {} {\bibfield  {journal} {\bibinfo  {journal} {J.
  Stat. Phys.}\ }\textbf {\bibinfo {volume} {168}},\ \bibinfo {pages}
  {937} (\bibinfo {year} {2017})}\BibitemShut {NoStop}%
\bibitem [{\citenamefont {{E. Akkermans and G.
  Montambaux}}(2007)}]{Akkermans:book07}%
  \BibitemOpen
  \bibfield  {author} {\bibinfo {author} {\bibnamefont {{E. Akkermans and G.
  Montambaux}}},\ }\href@noop {} {\emph {\bibinfo {title} {Mesoscopic Physics
  of Electrons and Photons}}}\ (\bibinfo  {publisher} {Cambridge University
  Press},\ \bibinfo {address} {Cambridge},\ \bibinfo {year} {2007})\BibitemShut
  {NoStop}%
\bibitem [{\citenamefont {Lee}\ \emph {et~al.}(1987)\citenamefont {Lee},
  \citenamefont {Stone},\ and\ \citenamefont {Fukuyama}}]{PhysRevB.35.1039}%
  \BibitemOpen
  \bibfield  {author} {\bibinfo {author} {\bibfnamefont {P.~A.}\ \bibnamefont
  {Lee}}, \bibinfo {author} {\bibfnamefont {A.~D.}\ \bibnamefont {Stone}}, \
  and\ \bibinfo {author} {\bibfnamefont {H.}~\bibnamefont {Fukuyama}},\ } {\bibfield  {journal} {\bibinfo
  {journal} {Phys. Rev. B}\ }\textbf {\bibinfo {volume} {35}},\ \bibinfo
  {pages} {1039} (\bibinfo {year} {1987})}\BibitemShut {NoStop}%
\bibitem [{\citenamefont {Altland}(1994)}]{altland1994electronic}%
  \BibitemOpen
  \bibfield  {author} {\bibinfo {author} {\bibfnamefont {A.}~\bibnamefont
  {Altland}},\ }\href@noop {} {\bibfield  {journal} {\bibinfo  {journal}
  {Ann. Phys. (Germany)}\ }\textbf {\bibinfo {volume} {506}},\ \bibinfo {pages}
  {28} (\bibinfo {year} {1994})}\BibitemShut {NoStop}%
\bibitem [{\citenamefont {Bergmann}(1984)}]{Bergmann:PR84}%
  \BibitemOpen
  \bibfield  {author} {\bibinfo {author} {\bibfnamefont {G.}~\bibnamefont
  {Bergmann}},\ }\href@noop {} {\bibfield  {journal} {\bibinfo  {journal}
  {Phys. Rep.}\ }\textbf {\bibinfo {volume} {107}},\ \bibinfo {pages} {1}
  (\bibinfo {year} {1984})}\BibitemShut {NoStop}%
\bibitem [{\citenamefont {Genack}\ and\ \citenamefont
  {Wang}(2010)}]{genack2010speckle}%
  \BibitemOpen
  \bibfield  {author} {\bibinfo {author} {\bibfnamefont {A.~Z.}\ \bibnamefont
  {Genack}}\ and\ \bibinfo {author} {\bibfnamefont {J.}~\bibnamefont {Wang}},\
  }\href@noop {} {\bibfield  {journal} {\bibinfo  {journal} {Int.
  J. Mod. Phys. B}\ }\textbf {\bibinfo {volume} {24}},\ \bibinfo
  {pages} {1950} (\bibinfo {year} {2010})}\BibitemShut {NoStop}%
\bibitem [{\citenamefont {Lagendijk}\ \emph {et~al.}(2009)\citenamefont
  {Lagendijk}, \citenamefont {van Tiggelen},\ and\ \citenamefont
  {Wiersma}}]{lagendijk2009fifty}%
  \BibitemOpen
  \bibfield  {author} {\bibinfo {author} {\bibfnamefont {A.}~\bibnamefont
  {Lagendijk}}, \bibinfo {author} {\bibfnamefont {B.}~\bibnamefont {van
  Tiggelen}}, \ and\ \bibinfo {author} {\bibfnamefont {D.}~\bibnamefont
  {Wiersma}},\ }\href@noop {} {\bibfield  {journal} {\bibinfo  {journal}
  {Physics Today}\ }\textbf {\bibinfo {volume} {62}},\ \bibinfo {pages} {24}
  (\bibinfo {year} {2009})}\BibitemShut {NoStop}%
\bibitem [{\citenamefont {Sanchez-Palencia}\ and\ \citenamefont
  {Lewenstein}(2010)}]{sanchez2010disordered}%
  \BibitemOpen
  \bibfield  {author} {\bibinfo {author} {\bibfnamefont {L.}~\bibnamefont
  {Sanchez-Palencia}}\ and\ \bibinfo {author} {\bibfnamefont {M.}~\bibnamefont
  {Lewenstein}},\ }\href@noop {} {\bibfield  {journal} {\bibinfo  {journal}
  {Nat. Phys.}\ }\textbf {\bibinfo {volume} {6}},\ \bibinfo {pages} {87}
  (\bibinfo {year} {2010})}\BibitemShut {NoStop}%
\bibitem [{\citenamefont {Aspect}\ and\ \citenamefont
  {Inguscio}(2009)}]{aspect2009anderson}%
  \BibitemOpen
  \bibfield  {author} {\bibinfo {author} {\bibfnamefont {A.}~\bibnamefont
  {Aspect}}\ and\ \bibinfo {author} {\bibfnamefont {M.}~\bibnamefont
  {Inguscio}},\ }\href@noop {} {\bibfield  {journal} {\bibinfo  {journal}
  {Physics Today}\ }\textbf {\bibinfo {volume} {62}},\ \bibinfo {pages} {30}
  (\bibinfo {year} {2009})}\BibitemShut {NoStop}%
\bibitem [{\citenamefont {Lopez}\ \emph {et~al.}(2013)\citenamefont {Lopez},
  \citenamefont {Cl{\'e}ment}, \citenamefont {{G. Lemari{\'e}}}, \citenamefont
  {Delande}, \citenamefont {Szriftgiser},\ and\ \citenamefont
  {Garreau}}]{lopez2013phase}%
  \BibitemOpen
  \bibfield  {author} {\bibinfo {author} {\bibfnamefont {M.}~\bibnamefont
  {Lopez}}, \bibinfo {author} {\bibfnamefont {J.-F.}\ \bibnamefont
  {Cl{\'e}ment}}, \bibinfo {author} {\bibnamefont {{G. Lemari{\'e}}}}, \bibinfo
  {author} {\bibfnamefont {D.}~\bibnamefont {Delande}}, \bibinfo {author}
  {\bibfnamefont {P.}~\bibnamefont {Szriftgiser}}, \ and\ \bibinfo {author}
  {\bibfnamefont {J.~C.}\ \bibnamefont {Garreau}},\ }\href@noop {} {\bibfield
  {journal} {\bibinfo  {journal} {New J. Phys.}\ }\textbf {\bibinfo {volume}
  {15}},\ \bibinfo {pages} {065013} (\bibinfo {year} {2013})}\BibitemShut
  {NoStop}%
\bibitem [{\citenamefont {Manai}\ \emph {et~al.}(2015)\citenamefont {Manai},
  \citenamefont {Cl\'ement}, \citenamefont {Chicireanu}, \citenamefont
  {Hainaut}, \citenamefont {Garreau}, \citenamefont {Szriftgiser},\ and\
  \citenamefont {Delande}}]{PhysRevLett.115.240603}%
  \BibitemOpen
  \bibfield  {author} {\bibinfo {author} {\bibfnamefont {I.}~\bibnamefont
  {Manai}}, \bibinfo {author} {\bibfnamefont {J.-F.}\ \bibnamefont
  {Cl\'ement}}, \bibinfo {author} {\bibfnamefont {R.}~\bibnamefont
  {Chicireanu}}, \bibinfo {author} {\bibfnamefont {C.}~\bibnamefont {Hainaut}},
  \bibinfo {author} {\bibfnamefont {J.~C.}\ \bibnamefont {Garreau}}, \bibinfo
  {author} {\bibfnamefont {P.}~\bibnamefont {Szriftgiser}}, \ and\ \bibinfo
  {author} {\bibfnamefont {D.}~\bibnamefont {Delande}},\ }
  {\bibfield  {journal} {\bibinfo  {journal}
  {Phys. Rev. Lett.}\ }\textbf {\bibinfo {volume} {115}},\ \bibinfo {pages}
  {240603} (\bibinfo {year} {2015})}\BibitemShut {NoStop}%
\bibitem [{\citenamefont {Krinner}\ \emph {et~al.}(2017)\citenamefont
  {Krinner}, \citenamefont {Esslinger},\ and\ \citenamefont
  {Brantut}}]{0953-8984-29-34-343003}%
  \BibitemOpen
  \bibfield  {author} {\bibinfo {author} {\bibfnamefont {S.}~\bibnamefont
  {Krinner}}, \bibinfo {author} {\bibfnamefont {T.}~\bibnamefont {Esslinger}},
  \ and\ \bibinfo {author} {\bibfnamefont {J.-P.}\ \bibnamefont {Brantut}},\
  } {\bibfield
  {journal} {\bibinfo  {journal} {J. Phys. Condens. Matter}\
  }\textbf {\bibinfo {volume} {29}},\ \bibinfo {pages} {343003} (\bibinfo
  {year} {2017})}\BibitemShut {NoStop}%
\bibitem [{\citenamefont {Krinner}\ \emph {et~al.}(2015)\citenamefont
  {Krinner}, \citenamefont {Stadler}, \citenamefont {Meineke}, \citenamefont
  {Brantut},\ and\ \citenamefont {Esslinger}}]{PhysRevLett.115.045302}%
  \BibitemOpen
  \bibfield  {author} {\bibinfo {author} {\bibfnamefont {S.}~\bibnamefont
  {Krinner}}, \bibinfo {author} {\bibfnamefont {D.}~\bibnamefont {Stadler}},
  \bibinfo {author} {\bibfnamefont {J.}~\bibnamefont {Meineke}}, \bibinfo
  {author} {\bibfnamefont {J.-P.}\ \bibnamefont {Brantut}}, \ and\ \bibinfo
  {author} {\bibfnamefont {T.}~\bibnamefont {Esslinger}},\ }
  {\bibfield  {journal} {\bibinfo  {journal}
  {Phys. Rev. Lett.}\ }\textbf {\bibinfo {volume} {115}},\ \bibinfo {pages}
  {045302} (\bibinfo {year} {2015})}\BibitemShut {NoStop}%
\bibitem [{\citenamefont {H\"ausler}\ \emph {et~al.}(2017)\citenamefont
  {H\"ausler}, \citenamefont {Nakajima}, \citenamefont {Lebrat}, \citenamefont
  {Husmann}, \citenamefont {Krinner}, \citenamefont {Esslinger},\ and\
  \citenamefont {Brantut}}]{PhysRevLett.119.030403}%
  \BibitemOpen
  \bibfield  {author} {\bibinfo {author} {\bibfnamefont {S.}~\bibnamefont
  {H\"ausler}}, \bibinfo {author} {\bibfnamefont {S.}~\bibnamefont {Nakajima}},
  \bibinfo {author} {\bibfnamefont {M.}~\bibnamefont {Lebrat}}, \bibinfo
  {author} {\bibfnamefont {D.}~\bibnamefont {Husmann}}, \bibinfo {author}
  {\bibfnamefont {S.}~\bibnamefont {Krinner}}, \bibinfo {author} {\bibfnamefont
  {T.}~\bibnamefont {Esslinger}}, \ and\ \bibinfo {author} {\bibfnamefont
  {J.-P.}\ \bibnamefont {Brantut}},\ } {\bibfield  {journal} {\bibinfo  {journal}
  {Phys. Rev. Lett.}\ }\textbf {\bibinfo {volume} {119}},\ \bibinfo {pages}
  {030403} (\bibinfo {year} {2017})}\BibitemShut {NoStop}%
\bibitem [{\citenamefont {Fisher}\ \emph {et~al.}(1989)\citenamefont {Fisher},
  \citenamefont {Weichman}, \citenamefont {Grinstein},\ and\ \citenamefont
  {Fisher}}]{Fisher:dirtyboson:PRB89}%
  \BibitemOpen
  \bibfield  {author} {\bibinfo {author} {\bibfnamefont {M.~P.~A.}\
  \bibnamefont {Fisher}}, \bibinfo {author} {\bibfnamefont {P.~B.}\
  \bibnamefont {Weichman}}, \bibinfo {author} {\bibfnamefont {G.}~\bibnamefont
  {Grinstein}}, \ and\ \bibinfo {author} {\bibfnamefont {D.~S.}\ \bibnamefont
  {Fisher}},\ } {\bibfield  {journal}
  {\bibinfo  {journal} {Phys. Rev. B}\ }\textbf {\bibinfo {volume} {40}},\
  \bibinfo {pages} {546} (\bibinfo {year} {1989})}\BibitemShut {NoStop}%
\bibitem [{\citenamefont {Giamarchi}\ and\ \citenamefont
  {Schulz}(1988)}]{Giamarchi1988a}%
  \BibitemOpen
  \bibfield  {author} {\bibinfo {author} {\bibfnamefont {T.}~\bibnamefont
  {Giamarchi}}\ and\ \bibinfo {author} {\bibfnamefont {H.~J.}\ \bibnamefont
  {Schulz}},\ } {\bibfield  {journal}
  {\bibinfo  {journal} {Phys. Rev. B}\ }\textbf {\bibinfo {volume} {37}},\
  \bibinfo {pages} {325} (\bibinfo {year} {1988})}\BibitemShut {NoStop}%
\bibitem [{\citenamefont {Altman}\ \emph {et~al.}(2008)\citenamefont {Altman},
  \citenamefont {Kafri}, \citenamefont {Polkovnikov},\ and\ \citenamefont
  {Refael}}]{Altman2008a}%
  \BibitemOpen
  \bibfield  {author} {\bibinfo {author} {\bibfnamefont {E.}~\bibnamefont
  {Altman}}, \bibinfo {author} {\bibfnamefont {Y.}~\bibnamefont {Kafri}},
  \bibinfo {author} {\bibfnamefont {A.}~\bibnamefont {Polkovnikov}}, \ and\
  \bibinfo {author} {\bibfnamefont {G.}~\bibnamefont {Refael}},\ }
  {\bibfield  {journal} {\bibinfo
  {journal} {Phys. Rev. Lett.}\ }\textbf {\bibinfo {volume} {100}},\ \bibinfo
  {pages} {170402} (\bibinfo {year} {2008})}\BibitemShut {NoStop}%
\bibitem [{\citenamefont {Doggen}\ \emph {et~al.}(2017)\citenamefont {Doggen},
  \citenamefont {Lemari\'e}, \citenamefont {Capponi},\ and\ \citenamefont
  {Laflorencie}}]{doggen2017weak}%
  \BibitemOpen
  \bibfield  {author} {\bibinfo {author} {\bibfnamefont {E.~V.~H.}\
  \bibnamefont {Doggen}}, \bibinfo {author} {\bibfnamefont {G.}~\bibnamefont
  {Lemari\'e}}, \bibinfo {author} {\bibfnamefont {S.}~\bibnamefont {Capponi}},
  \ and\ \bibinfo {author} {\bibfnamefont {N.}~\bibnamefont {Laflorencie}},\
  } {\bibfield  {journal} {\bibinfo
  {journal} {Phys. Rev. B}\ }\textbf {\bibinfo {volume} {96}},\ \bibinfo
  {pages} {180202} (\bibinfo {year} {2017})}\BibitemShut {NoStop}%
\bibitem [{\citenamefont {M{\"u}ller}(2013)}]{muller2013magnetoresistance}%
  \BibitemOpen
  \bibfield  {author} {\bibinfo {author} {\bibfnamefont {M.}~\bibnamefont
  {M{\"u}ller}},\ }\href@noop {} {\bibfield  {journal} {\bibinfo  {journal}
  {Europhys. Lett.}\ }\textbf {\bibinfo {volume} {102}},\ \bibinfo
  {pages} {67008} (\bibinfo {year} {2013})}\BibitemShut {NoStop}%
\bibitem [{\citenamefont {Gangopadhyay}\ \emph {et~al.}(2013)\citenamefont
  {Gangopadhyay}, \citenamefont {Galitski},\ and\ \citenamefont
  {M{\"u}ller}}]{gangopadhyay2013magnetoresistance}%
  \BibitemOpen
  \bibfield  {author} {\bibinfo {author} {\bibfnamefont {A.}~\bibnamefont
  {Gangopadhyay}}, \bibinfo {author} {\bibfnamefont {V.}~\bibnamefont
  {Galitski}}, \ and\ \bibinfo {author} {\bibfnamefont {M.}~\bibnamefont
  {M{\"u}ller}},\ }\href@noop {} {\bibfield  {journal} {\bibinfo  {journal}
  {Phys. Rev. Lett.}\ }\textbf {\bibinfo {volume} {111}},\ \bibinfo
  {pages} {026801} (\bibinfo {year} {2013})}\BibitemShut {NoStop}%
\bibitem [{\citenamefont {Monthus}\ and\ \citenamefont
  {Garel}(2012)}]{monthus2012random}%
  \BibitemOpen
  \bibfield  {author} {\bibinfo {author} {\bibfnamefont {C.}~\bibnamefont
  {Monthus}}\ and\ \bibinfo {author} {\bibfnamefont {T.}~\bibnamefont
  {Garel}},\ }\href@noop {} {\bibfield  {journal} {\bibinfo  {journal} {J.
  Stat. Mech.}\ }\textbf {\bibinfo {volume}
  {2012}},\ \bibinfo {pages} {P01008} (\bibinfo {year} {2012})}\BibitemShut
  {NoStop}%
\end{thebibliography}

\begin{thebibliography}{14}%
\makeatletter
\providecommand \@ifxundefined [1]{%
 \@ifx{#1\undefined}
}%
\providecommand \@ifnum [1]{%
 \ifnum #1\expandafter \@firstoftwo
 \else \expandafter \@secondoftwo
 \fi
}%
\providecommand \@ifx [1]{%
 \ifx #1\expandafter \@firstoftwo
 \else \expandafter \@secondoftwo
 \fi
}%
\providecommand \natexlab [1]{#1}%
\providecommand \enquote  [1]{``#1''}%
\providecommand \bibnamefont  [1]{#1}%
\providecommand \bibfnamefont [1]{#1}%
\providecommand \citenamefont [1]{#1}%
\providecommand \href@noop [0]{\@secondoftwo}%
\providecommand \href [0]{\begingroup \@sanitize@url \@href}%
\providecommand \@href[1]{\@@startlink{#1}\@@href}%
\providecommand \@@href[1]{\endgroup#1\@@endlink}%
\providecommand \@sanitize@url [0]{\catcode `\\12\catcode `\$12\catcode
  `\&12\catcode `\#12\catcode `\^12\catcode `\_12\catcode `\%12\relax}%
\providecommand \@@startlink[1]{}%
\providecommand \@@endlink[0]{}%
\providecommand \url  [0]{\begingroup\@sanitize@url \@url }%
\providecommand \@url [1]{\endgroup\@href {#1}{\urlprefix }}%
\providecommand \urlprefix  [0]{URL }%
\providecommand \Eprint [0]{\href }%
\providecommand \doibase [0]{http://dx.doi.org/}%
\providecommand \selectlanguage [0]{\@gobble}%
\providecommand \bibinfo  [0]{\@secondoftwo}%
\providecommand \bibfield  [0]{\@secondoftwo}%
\providecommand \translation [1]{[#1]}%
\providecommand \BibitemOpen [0]{}%
\providecommand \bibitemStop [0]{}%
\providecommand \bibitemNoStop [0]{.\EOS\space}%
\providecommand \EOS [0]{\spacefactor3000\relax}%
\providecommand \BibitemShut  [1]{\csname bibitem#1\endcsname}%
\let\auto@bib@innerbib\@empty
%</preamble>
\bibitem [{\citenamefont {Somoza}\ \emph {et~al.}(2007)\citenamefont {Somoza},
  \citenamefont {Ortu\~no},\ and\ \citenamefont
  {Prior}}]{PhysRevLett.99.116602}%
  \BibitemOpen
  \bibfield  {author} {\bibinfo {author} {\bibfnamefont {A.~M.}\ \bibnamefont
  {Somoza}}, \bibinfo {author} {\bibfnamefont {M.}~\bibnamefont {Ortu\~no}}, \
  and\ \bibinfo {author} {\bibfnamefont {J.}~\bibnamefont {Prior}},\ }{\bibfield  {journal} {\bibinfo
  {journal} {Phys. Rev. Lett.}\ }\textbf {\bibinfo {volume} {99}},\ \bibinfo
  {pages} {116602} (\bibinfo {year} {2007})}\BibitemShut {NoStop}%
\bibitem [{\citenamefont {Prior}\ \emph {et~al.}(2005)\citenamefont {Prior},
  \citenamefont {Somoza},\ and\ \citenamefont {Ortuno}}]{prior2005conductance}%
  \BibitemOpen
  \bibfield  {author} {\bibinfo {author} {\bibfnamefont {J.}~\bibnamefont
  {Prior}}, \bibinfo {author} {\bibfnamefont {A.}~\bibnamefont {Somoza}}, \
  and\ \bibinfo {author} {\bibfnamefont {M.}~\bibnamefont {Ortuno}},\
  }\href@noop {} {\bibfield  {journal} {\bibinfo  {journal} {Physical Review
  B}\ }\textbf {\bibinfo {volume} {72}},\ \bibinfo {pages} {024206} (\bibinfo
  {year} {2005})}\BibitemShut {NoStop}%
\bibitem [{\citenamefont {Prior}\ \emph {et~al.}(2009)\citenamefont {Prior},
  \citenamefont {Somoza},\ and\ \citenamefont {Ortuno}}]{prior2009conductance}%
  \BibitemOpen
  \bibfield  {author} {\bibinfo {author} {\bibfnamefont {J.}~\bibnamefont
  {Prior}}, \bibinfo {author} {\bibfnamefont {A.}~\bibnamefont {Somoza}}, \
  and\ \bibinfo {author} {\bibfnamefont {M.}~\bibnamefont {Ortuno}},\
  }\href@noop {} {\bibfield  {journal} {\bibinfo  {journal} {The European
  Physical Journal B}\ }\textbf {\bibinfo {volume} {70}},\ \bibinfo {pages}
  {513} (\bibinfo {year} {2009})}\BibitemShut {NoStop}%
\bibitem [{\citenamefont {Halpin-Healy}\ and\ \citenamefont
  {Zhang}(1995)}]{halpin1995kinetic}%
  \BibitemOpen
  \bibfield  {author} {\bibinfo {author} {\bibfnamefont {T.}~\bibnamefont
  {Halpin-Healy}}\ and\ \bibinfo {author} {\bibfnamefont {Y.-C.}\ \bibnamefont
  {Zhang}},\ }\href@noop {} {\bibfield  {journal} {\bibinfo  {journal} {Phys.
  Rep.}\ }\textbf {\bibinfo {volume} {254}},\ \bibinfo {pages} {215}
  (\bibinfo {year} {1995})}\BibitemShut {NoStop}%
\bibitem [{\citenamefont {Johansson}(2000)}]{johansson2000shape}%
  \BibitemOpen
  \bibfield  {author} {\bibinfo {author} {\bibfnamefont {K.}~\bibnamefont
  {Johansson}},\ }\href@noop {} {\bibfield  {journal} {\bibinfo  {journal}
  {Communications in mathematical physics}\ }\textbf {\bibinfo {volume}
  {209}},\ \bibinfo {pages} {437} (\bibinfo {year} {2000})}\BibitemShut
  {NoStop}%
\bibitem [{\citenamefont {Tracy}\ and\ \citenamefont
  {Widom}(1994)}]{tracy1994level}%
  \BibitemOpen
  \bibfield  {author} {\bibinfo {author} {\bibfnamefont {C.~A.}\ \bibnamefont
  {Tracy}}\ and\ \bibinfo {author} {\bibfnamefont {H.}~\bibnamefont {Widom}},\
  }\href@noop {} {\bibfield  {journal} {\bibinfo  {journal} {Commun. 
  Math. Phys.}\ }\textbf {\bibinfo {volume} {159}},\ \bibinfo {pages}
  {151} (\bibinfo {year} {1994})}\BibitemShut {NoStop}%
\bibitem [{\citenamefont {Tracy}\ and\ \citenamefont
  {Widom}(1996)}]{tracy1996orthogonal}%
  \BibitemOpen
  \bibfield  {author} {\bibinfo {author} {\bibfnamefont {C.~A.}\ \bibnamefont
  {Tracy}}\ and\ \bibinfo {author} {\bibfnamefont {H.}~\bibnamefont {Widom}},\
  }\href@noop {} {\bibfield  {journal} {\bibinfo  {journal} {Commun.
  Math. Phys.}\ }\textbf {\bibinfo {volume} {177}},\ \bibinfo {pages}
  {727} (\bibinfo {year} {1996})}\BibitemShut {NoStop}%
\bibitem [{\citenamefont {Somoza}\ \emph {et~al.}(2015)\citenamefont {Somoza},
  \citenamefont {Le~Doussal},\ and\ \citenamefont
  {Ortu\~no}}]{PhysRevB.91.155413}%
  \BibitemOpen
  \bibfield  {author} {\bibinfo {author} {\bibfnamefont {A.~M.}\ \bibnamefont
  {Somoza}}, \bibinfo {author} {\bibfnamefont {P.}~\bibnamefont {Le~Doussal}},
  \ and\ \bibinfo {author} {\bibfnamefont {M.}~\bibnamefont {Ortu\~no}},\
  } {\bibfield  {journal} {\bibinfo
  {journal} {Phys. Rev. B}\ }\textbf {\bibinfo {volume} {91}},\ \bibinfo
  {pages} {155413} (\bibinfo {year} {2015})}\BibitemShut {NoStop}%
\bibitem [{\citenamefont {Groth}\ \emph {et~al.}(2014)\citenamefont {Groth},
  \citenamefont {Wimmer}, \citenamefont {Akhmerov},\ and\ \citenamefont
  {Waintal}}]{groth2014kwant}%
  \BibitemOpen
  \bibfield  {author} {\bibinfo {author} {\bibfnamefont {C.~W.}\ \bibnamefont
  {Groth}}, \bibinfo {author} {\bibfnamefont {M.}~\bibnamefont {Wimmer}},
  \bibinfo {author} {\bibfnamefont {A.~R.}\ \bibnamefont {Akhmerov}}, \ and\
  \bibinfo {author} {\bibfnamefont {X.}~\bibnamefont {Waintal}},\ }\href@noop
  {} {\bibfield  {journal} {\bibinfo  {journal} {N. J. Phys.}\
  }\textbf {\bibinfo {volume} {16}},\ \bibinfo {pages} {063065} (\bibinfo
  {year} {2014})}\BibitemShut {NoStop}%
\bibitem [{\citenamefont {Marko{\v{s}}}(2010)}]{markovs2010electron}%
  \BibitemOpen
  \bibfield  {author} {\bibinfo {author} {\bibfnamefont {P.}~\bibnamefont
  {Marko{\v{s}}}},\ }\href@noop {} {\bibfield  {journal} {\bibinfo  {journal}
  {Physica B}\ }\textbf {\bibinfo {volume} {405}},\ \bibinfo
  {pages} {3029} (\bibinfo {year} {2010})}\BibitemShut {NoStop}%
\bibitem [{\citenamefont {Maes}\ and\ \citenamefont
  {Thiery}(2017)}]{maes2017midpoint}%
  \BibitemOpen
  \bibfield  {author} {\bibinfo {author} {\bibfnamefont {C.}~\bibnamefont
  {Maes}}\ and\ \bibinfo {author} {\bibfnamefont {T.}~\bibnamefont {Thiery}},\
  }\href@noop {} {\bibfield  {journal} {\bibinfo  {journal} {J.
  Stat. Phys.}\ }\textbf {\bibinfo {volume} {168}},\ \bibinfo {pages}
  {937} (\bibinfo {year} {2017})}\BibitemShut {NoStop}%
\bibitem [{\citenamefont {Gorini}\ \emph {et~al.}(2013)\citenamefont {Gorini},
  \citenamefont {Jalabert}, \citenamefont {Szewc}, \citenamefont {Tomsovic},\
  and\ \citenamefont {Weinmann}}]{gorini2013theory}%
  \BibitemOpen
  \bibfield  {author} {\bibinfo {author} {\bibfnamefont {C.}~\bibnamefont
  {Gorini}}, \bibinfo {author} {\bibfnamefont {R.~A.}\ \bibnamefont
  {Jalabert}}, \bibinfo {author} {\bibfnamefont {W.}~\bibnamefont {Szewc}},
  \bibinfo {author} {\bibfnamefont {S.}~\bibnamefont {Tomsovic}}, \ and\
  \bibinfo {author} {\bibfnamefont {D.}~\bibnamefont {Weinmann}},\ }\href@noop
  {} {\bibfield  {journal} {\bibinfo  {journal} {Phys. Rev. B}\ }\textbf
  {\bibinfo {volume} {88}},\ \bibinfo {pages} {035406} (\bibinfo {year}
  {2013})}\BibitemShut {NoStop}%
\bibitem [{\citenamefont {Abbout}\ \emph {et~al.}(2011)\citenamefont {Abbout},
  \citenamefont {Lemari{\'e}},\ and\ \citenamefont
  {Pichard}}]{abbout2011thermal}%
  \BibitemOpen
  \bibfield  {author} {\bibinfo {author} {\bibfnamefont {A.}~\bibnamefont
  {Abbout}}, \bibinfo {author} {\bibfnamefont {G.}~\bibnamefont {Lemari{\'e}}},
  \ and\ \bibinfo {author} {\bibfnamefont {J.-L.}\ \bibnamefont {Pichard}},\
  }\href@noop {} {\bibfield  {journal} {\bibinfo  {journal} {Phys. Rev.
  Lett.}\ }\textbf {\bibinfo {volume} {106}},\ \bibinfo {pages} {156810}
  (\bibinfo {year} {2011})}\BibitemShut {NoStop}%
\bibitem [{\citenamefont {Wilhelm}\ \emph {et~al.}(2015)\citenamefont
  {Wilhelm}, \citenamefont {Walz},\ and\ \citenamefont
  {Evers}}]{wilhelm2015ab}%
  \BibitemOpen
  \bibfield  {author} {\bibinfo {author} {\bibfnamefont {J.}~\bibnamefont
  {Wilhelm}}, \bibinfo {author} {\bibfnamefont {M.}~\bibnamefont {Walz}}, \
  and\ \bibinfo {author} {\bibfnamefont {F.}~\bibnamefont {Evers}},\
  }\href@noop {} {\bibfield  {journal} {\bibinfo  {journal} {Phys. Rev.
  B}\ }\textbf {\bibinfo {volume} {92}},\ \bibinfo {pages} {014405} (\bibinfo
  {year} {2015})}\BibitemShut {NoStop}%
\end{thebibliography}
\end {document}